**Divergence vs. Decision P-values: A Distinction Worth Making in Theory and Keeping in Practice – or, How Divergence P-values Measure Evidence Even When Decision P-values Do Not**

SANDER GREENLAND

*Department of Epidemiology and Department of Statistics, University of California, Los Angeles, California, U.S.A., lesdomes@ucla.edu*

**Incorporates additions and corrections to print version (Greenland, 2023a) up to 12 September 2023, shown in bold on p. 5 and 23-24. See also comments and rejoinder (Greenland, 2023b).**

ABSTRACT**.** There are two distinct definitions of "P-value" for evaluating a proposed hypothesis or model for the process generating an observed dataset. The original definition starts with a measure of the divergence of the dataset from what was expected under the model, such as a sum of squares or a deviance statistic. A P-value is then the ordinal location of the measure in a reference distribution computed from the model and the data, and is treated as a unit-scaled index of compatibility between the data and the model. In the other definition, a P-value is a random variable on the unit interval whose realizations can be compared to a cutoff α to generate a decision rule with known error rates under the model and specific alternatives. It is commonly assumed that realizations of such decision P-values always correspond to divergence P-values. But this need not be so: Decision P-values can violate intuitive single-sample coherence criteria where divergence P-values do not. It is thus argued that divergence and decision P-values should be carefully distinguished in teaching, and that divergence P-values are the relevant choice when the analysis goal is to summarize evidence rather than implement a decision rule.

**Introduction: Two nonequivalent concepts of P-values**

There are at least two distinct definitions and uses of P-values within frequentist statistics. The first historically is the divergence P-value, which in its basic form is a descriptive geometric summary of the *observed* deviation of a discrepancy statistic from a model prediction. The second is the decision P-value, which is either a random variable used for compact expression of discrete-choice rules over an entire sample space, or the observed value of such a random variable.

In simple problems divergence P-values equal realizations of decision P-values; hence the two types are usually not distinguished. The present paper explores how they nonetheless do not correspond in general, either mathematically or in meaning. It also provides some theoretical development for divergence P-values to explain in detail why criticisms of P-values as evidence measures apply to decision P-values but not to divergence P-values. In particular, claims that P-values are incoherent evidence measures arise from the use of uniformly most powerful unbiasedness (UMPU) as a criterion to select decision rules, and use of the resulting decision P-values to measure support of hypotheses or models. The claims do not apply when instead P-values are selected to satisfy single-sample coherence properties and used to measure compatibility, consonance, or consistency of data with hypotheses or models.

The distinction is not new – see for example Kempthorne (1976), and also Cox (1977) in this journal over 45 years ago. The present paper elaborates the distinction in both mathematical and philosophical terms to respond to attacks on the use of P-values as evidence measures, which are based on a confusion of divergence and decision P-values, and a confusion of compatibility with support. As an extreme example, one recent article entitled "P-values Don't Measure Evidence" claimed that all descriptions of P-values as evidence measures are erroneous (Lavine, 2022). The examples used to argue this claim are however based on decision-theoretic P-values and likelihood-based criteria for evidence, and overlook entirely definitions and criteria aimed instead at summarizing geometric fit of observations to models. In the latter definitions, P-values simply describe the divergences between projections of data onto models for sampling probabilities. A P-value is then seen as one measure of one limited dimension of evidence against a statistical hypothesis or model, although it better serves this role if log-transformed to a more equal-interval scale.

Lavine's claim should be contrasted to (for example) Casella & Berger (1987a, p. 106), who state "to a frequentist, **evidence takes the form of the p value**, or the observed level of significance of the result", and to the writings of Karl Pearson, R.A. Fisher, D.R. Cox, and many others who used P-values as part of descriptions of *refutational* evidence measures (evidence against statistical hypotheses). Here are some examples with added emphases; note that 20th-century British statisticians often used "level of significance" and "value of P" to refer to what American statisticians came to call P-values, the latter term having appeared by the 1920s (see Shafer, 2020):

> "…the larger the value of [the observed test statistic] t the stronger the **evidence of departure from** $H_0$ of the type it is required to test… For given observations y we calculate t = $t_{obs}$ = t(y), say, and the *level of significance* $p_{obs}$ by $p_{obs}$ = pr(T≥ $t_{obs}$;$H_0$)… We use $p_{obs}$ as a **measure of the consistency** of the data with $H_0$ with the following hypothetical interpretation: Suppose that we were to accept the available data as **evidence against** $H_0$. Then we would be bound to accept all data with a larger value of t as even stronger **evidence**." – Cox & Hinkley, *Theoretical Statistics* (1974, p. 66).

> "…if the value of v [the test statistic] is in the lower or central part of the distribution [of v under $H_0$], the data are **consistent with** $H_0$; or if v is in the extreme upper tail of the distribution then this is **evidence against** $H_0$…the p-value is in a sense the unique measure of the extremity of the value of v in light of $H_0$." – Cox & Donnelly (2011, p. 146-147).

In these sources the observed P-value is a strictly monotone transform of the test statistic via the latter's cumulative distribution function under $H_0$, which makes $p_{obs}$ the quantile at which $t_{obs}$ fell (Perezgonzalez, 2015a). Thus, in the above quotes the random statistic T orders possible samples by the amount of evidence against $H_0$, and $p_{obs}$ translates $t_{obs}$ to the observed sample rank in that evidence ordering. Notably, a large value of $p_{obs}$ indicates the observations provide little evidence against $H_0$, but cannot by itself constitute evidence *for* $H_0$; it can only be taken as indicating consistency, compatibility, concordance, or consonance of the data with $H_0$.

Discrepancy P-values for evidence summarization have been traditionally discussed under the heading of pure or absolute significance tests, and labeled "observed significance levels" (Cox & Hinkley, 1974, Ch. 3; Cox, 1977, sec. 2.7). Cox & Hinkley (1974, p. 66) did invite confusion of evidential and decision P-values by saying that "$p_{obs}$ is the probability that we would mistakenly declare there to be **evidence** against $H_0$, were we to regard the data under analysis as **just decisive against** $H_0$." Subsequently however Cox (1977, sec. 2.1) noted that "such a procedure is to be distinguished sharply from a decision problem in which 'acceptance' or 'rejection' is required", and in sec. 2.3 emphasized the distinction in saying that "The contrast between significance tests, as an aid in the summarization of **evidence**, and decision procedures is implicit or explicit in much recent discussion."

Cox (1977, sec. 2.3) further explained that such evidential P-values were applicable in the absence of explicit alternatives to the tested constraint (hypothesis):

> "At this stage in the development [of a statistical model], the null hypothesis is the only aspect of the problem explicitly formulated, **so that such general considerations as the likelihood principle are inapplicable**."

Regarding the last bolded statement, Cox & Hinkley (1974, example 2.42 and exercise 2.16) give illustrations in which the likelihood function fails to capture evidence in a sensible or complete manner. More recent examples in which approaches obeying the strong likelihood principle (Cox & Hinkley, Ch. 2) fail to meet basic repeated-sampling (frequency-calibration) criteria, yet which admit valid P-values, are given by Robins & Ritov (1997), Robins & Wasserman (2000, sec. 5), and Ritov et al. (2014). Among philosophical criticisms of likelihood for capturing all aspects of evidence are Hacking (1980) and Mayo (2018, sec. 1.4). Thus, criticisms of P-values derived from violations of the likelihood principle are themselves disputable.

None of these failings of relative likelihood rule it out as *one* evidence summary to use among others. Rather, they should remind us that there is no single accepted definition of "statistical evidence", nor is there a single formalization of evidence concept that is flawless or adequate for all purposes. This fact should be no surprise: How could one justify restriction of a complex concept like "evidence" to one formalization when there is no single sufficient definition or measure of simpler concepts? Consider clothing size: To buy pants one needs at least waist size and leg length. Even some fundamental concepts in our most exact of sciences, physics, require multiple dimensions to capture (e.g., kinetic vs. potential energy). As Casella & Berger (1987b, p. 135) wrote,

> "Bayesians and frequentists may never agree on the appropriate way to analyze data and interpret results, but there is no reason why they cannot learn from one another. Whether or not measures of evidence can be reconciled is probably a minor consideration; **understanding what affects a measure of evidence is a major consideration**".

In line with the bolded statement, the present paper uses the P-value controversy to illustrate the split between reference (neoFisherian) frequentism, which focuses on describing relations of sampling distributions to observed data, and decision-theoretic (Neyman-Pearson-Wald) frequentism, which focuses on optimizing decisions over a sample space given a family of distributions on that space.

This split is usually obscured in teaching and research reports, despite attempts to bring to the fore (Goodman 1993, 2016; Hubbard & Bayarri 2003; Schneider, 2014; Perezgonzalez, 2015b). It is mirrored in the division between reference ("objective") Bayesianism, which focuses on summarizing parameter information in a data set, and "subjective" or operational Bayesianism, which focuses on optimizing decisions given external ("prior") information and, again, a family of distributions on the sample space. Formal decisions are however suspect if not misleading when there are important sources of uncertainty about the models on which they are based (Leamer, 1978; Stark, 2022). In contrast, divergence P-values can remain useful pieces of information for evaluating models under those conditions. The present treatment may in fact be seen as alternative generalization of the original concept of nondirectional P-values for model fit (Pearson, 1900), with the common 2-sided P-value for point hypotheses being a special case.

**In what follows I will use the terms "descriptive" and "discrepancy" to refer to P-values that can be interpreted as single-sample summaries of relations of models to data or to one another. That includes for example P-values from rank-sum statistics, which ignore spacing (metric) information on individuals when measuring discrepancies between distributions. I will use "divergence" more narrowly, to refer to descriptive P-values derivable from geometric-divergence statistics which employ that information (Amari, 2016); those include everyday statistics in parametric and semi-parametric modeling such as sum-of-squared-deviation and likelihood-ratio statistics. A divergence P-value is then an ordinal description of a geometric measurement.**

**Theoretical Background**

The present section describes the central arguments and findings in terms of general distribution-function spaces and may be skipped by less mathematically inclined readers. Subsequent sections will repeat its essential elements in less technical form.

Suppose the observed data y is idealized as an n-dimensional object that is a realization of random vector Y with unknown probability mass function f(y) on a sample space $R_U$ defined by the logically possible range of Y; usually Y is a real n-vector variable and hence $R_U \subseteq R^n$. This f(y) may be a composite of a density and a discrete mass function on $R_U$, with $f(y)$ denoting the density at continuity points of the distribution and the probability for mass points (jumps); $f(R_A)$ will also be used to denote the probability $\int_{R(A)} f(y)dy$ of $Y \varepsilon\ R(A) = R_A$ for an arbitrary subset $R_A$ of $R_U$.

Let $F_U$ be the unrestricted or "saturated" space of all possible probability mass functions on $R_U$, including single-point mass (degenerate) functions. Consider the case of evaluating a set A of constraints on or assumptions about the distribution of data generated by an observation process. Ideally, A is derived from a focused research hypothesis about the actual causal processes generating observations (Greenland, 2022). Taken together, the constraints in A may not fully specify f, but will define a set or family $F_A \subseteq F_U$ of the functions that satisfy all the constraints entailed by A. The complement of $F_A$ in $F_U$ is the set difference $F_U - F_A$, which will be denoted $\neg F_A$. Addition of further constraint sets H (traditional "null hypotheses") beyond A will be considered below. In model explorations, the constraint set A is sometimes called a working or embedding model for f.

*Discrepancy and Divergence P-values*

In its most basic form, a discrepancy P-value is an ordinal description of a real-valued summary measure $d(y;F_A)$ of the deviation or divergence of y from the set $\{\mu_f: f\varepsilon F_A\}$ of Y-expectations $\mu_f = E(Y;f)$ of the functions in $F_A$, where the expectations are assumed to be finite (the use of a semicolon rather than a comma or vertical bar within function arguments is to emphasize the asymmetry of the inputs without suggesting probabilistic conditioning, as in Cox and Hinkley, 1974 and Cox, 1977). Examples of such $d(y;F_A)$ include Pearson's $\chi^2$ statistic and the sum of squared standardized residuals in a regression analysis. The constraints in A are then used to derive a "reference" mass function $g_{y;A}(d) = g_A(d;y)$ for $D_A = d(Y;F_A)$ which may partially depend on y (as in conditional tests). To allow for distributions without expectations we may start from a general measure $d(f_1;f_0)$ of deviation of $f_1$ from $f_0$ in $F_U$ and define $d(y;F_A) = \inf\{d(f_y;f): f\varepsilon F_A\}$ where $f_y$ places all mass on y, i.e., $f_y(y) = 1$, so that $E(y;f_y) = y$. In all the settings considered here, $g_{y;A}$ is derived from $f_{y;A} = \arg\inf\{d(f_y;f): f\varepsilon F_A\}$ when the latter exists.

Assuming $D_A$ is a real scalar random variable such as a $\chi^2$ statistic, the observed upper-tail P-value derived from the realization $D_A = d$ is $p_{d;A} = g_{y;A}(\{D_A \geq d\}) = \int_{u \geq d} g_{y;A}(u)du$, which is the ordinal location or rank of d in the reference distribution. It is interpreted as a unit-scaled index of consonance, consistency or compatibility of y with $F_A$ (Kempthorne & Folks, 1971; Cox & Hinkley, 1974, p. 66; Cox, 1977; Box, 1980, p. 347; Folks, 1981; Bayarri & Berger, 2000; Robins et al., 2000; Greenland, 2019a; Amrhein et al., 2019; Rafi & Greenland, 2020; Amrhein & Greenland 2022; Greenland et al., 2022, 2023). In a complementary fashion, 1−$p_{d;A}$ is a unit-scaled index of incompatibility or inconsistency which summarizes the divergence of y from $F_A$ as measured by $d(y;F_A)$, although a more equal-interval measurement of incompatibility or refutational evidence is provided by the S-value −$\log_c(p_{d;A})$ where c is the log base (Kempthorne, 1990; Greenland, 2019a, 2021; Rafi & Greenland, 2020; Greenland et al., 2022, 2023). These definitions and interpretations are sometimes called Fisherian or neoFisherian, although they can be seen in writings before Fisher's, such as Pearson (1900), and could be called evidential or reference frequentist.

Note that, contrary to some discussions (e.g., Cox & Hinkley, 1974, p. 66) there is no requirement or implication that the random variable $P_{D;A} = \int_{u \geq D} g_{y;A}(u)du$ must approach uniformity when f$\varepsilon F_A$; in particular, $P_{D;A}$ will instead have a mass at 1 when $g_A$ has a mass at 0, as it will in the examples below. Nonetheless, under mild conditions $P_{D;A}$ will be distributed as or exceed a unit-uniform random variable when f$\varepsilon F_A$ and will have sampling properties that are intuitively desirable for a compatibility measure. For example, as n increases $P_{D;A}$ will concentrate toward 1 if the actual f has $d(f;\neg F_A)>0$, i.e., is interior to $F_A$; conversely, $P_{D;A}$ will concentrate toward 0 if the actual f has $d(f;F_A)>0$, i.e., is interior to $\neg F_A$.

Repeated-sampling properties are however insufficient to ensure single-sample coherency. The discrepancy $d(y;F_A)$ measures a departure of the single sample y from what the constraints A would lead us to expect. Here are two single-sample criteria we could require for using discrepancy P-values like $p_{d;A}$ (or strictly monotone transforms of them) as evidence measures:

1) Suppose the data exhibit no discrepancy whatsoever from the constraints in A, in the sense that $f_y \in F_A$; then we should want $d(y;F_A) = 0$ and hence $p_{d;A} = 1$.
2) Suppose M entails all constraints imposed by A and possibly more, so that $F_M \subseteq F_A$; then we should want $d(y;F_M) \geq d(y;F_A)$, with $p_{d;M} \leq p_{d;A}$ as well.

e.g., see Kempthorne (1976, p. 767). Suppose $d(y;F_A)$ is derived from a formal divergence measure $d(f_1;f_0)$ on $F_U \times F_U$ (Amari, 2016); condition (1) then follows immediately from the nonegativity of divergences. For condition (2), which may be termed subset coherence, suppose that $F_A$ is a subset of an exponential family of distributions (e.g., Y may be a vector of binomial variates). Denoting mean vectors by $\mu_k = E(Y;f_k)$ and covariance matrices by $\Sigma_k$, k=1,0, suppose $d(f_1;f_0) = (\mu_1-\mu_0)'\Sigma_0^{-1}(\mu_1-\mu_0)$, the squared $f_0$-standardized Euclidean distance between the Y-means of $f_1$ and $f_0$, sometime called the Euclidean divergence (although that label is sometimes applied to half this function; see Amari, 2016). We then have the divergence of the data from $F_A$ is $d(y;F_A) = \inf\{(y-\mu_0)'\Sigma_0^{-1}(y-\mu_0): f_0 \varepsilon F_A\}$, the Pearson $\chi^2$ fit statistic, while the mean $\mu_{y;A}$ of its minimizer $f_{y;A}$ is the minimum-$\chi^2$ or generalized-least-squares (GLS) estimate of $\mu$. Subset coherence then follows follows exactly when the covariance matrix $\Sigma$ is constant over $F_A$, with no need for approximate normality of Y, and follows in a local asymptotic sense given conventional regularity conditions.

Suppose instead we use the divergence of $F_A$ from the data, $d(F_A;y) = \inf\{d(f;f_y): f \varepsilon F_A\}$. Subset coherence then follows when taking $d(f_1;f_0) = E(\ln(f_0(Y))-\ln(f_1(Y));f_0)$, the Kullback-Leibler divergence (KLD) of $f_1$ from $f_0$, also known as the relative entropy or the discrimination information. The KLD is defined under the usual conventions that $0 \cdot \ln(0) = 0$ derived from $\lim(x \cdot \ln(x): x \downarrow 0) = 0$, and that $f_0(y) = 0$ when $f_1(y) = 0$ (absolute continuity of $f_0$ with respect to $f_1$). Because $f_y(y) = 1$ and is 0 elsewhere, we have $E(\ln(f_y(Y));f_y) = 0$, $E(\ln(f_1(Y));f_Y) = \ln(f_1(y))$ and thus $d(F_A;y) = \sup\{-\ln(f(y)): f \varepsilon F_A\}$, half the deviance statistic used in global tests of fit for regular GLMs. When it exists, the maximum-likelihood estimate (MLE) $\max\{f(y): f \varepsilon F_A\}$ of $f$ under A is then $f_{A;y} = \arg\inf\{d(F_A;y): f \varepsilon F_A\}$, and $f_y$ is the MLE of $f$ when A entails no constraint on $f$, as when $A = \varnothing$ (so that $F_A = F_U$).

*Decision P-values*

In the second definition, a P-value is merely a by-product of a decision criterion ("test") for whether to reject or accept the family $F_A$ for further use. The analyst specifies in advance a maximum acceptable false-rejection (Type-I error) rate $\alpha$ for $F_A$ and an alternative distribution family $F_{alt} \subset F_U$ disjoint from $F_A$. Where possible one then finds a critical region $R_A(\alpha) \subset R_U$ to implement the decision rule "reject $F_A$ if $y \varepsilon R_A(\alpha)$", for which

1) The type-I error rate is controlled at level $\alpha$ (test validity): When $f \varepsilon F_A$, $f(R_A(\alpha)) \leq \alpha$; i.e., when A holds, the probability $f(R_A(\alpha))$ of $Y \varepsilon R_A(\alpha)$ does not exceed $\alpha$.

2) The type-II error rate over $F_{alt}$ is minimized among valid tests: When $f \varepsilon F_{alt}$, $f(R_A(\alpha)) \geq f(R)$ for any other $R \subseteq R_U$ that has $f(R) \leq \alpha$ whenever $f \varepsilon F_A$; i.e., among valid tests (1), when A fails, $R_A(\alpha)$ maximizes the rejection probability (power).

3) The type-II error rate never falls below $\alpha$ (unbiasedness): When $f \varepsilon F_{alt}$, $f(R_A(\alpha)) \geq \alpha$.

If a set function $R_A(\alpha)$ of $\alpha$ satisfies 1-3 for all $\alpha$, the decision rule "reject $F_A$ if $y \varepsilon R_A(\alpha)$" is called a uniformly most powerful unbiased (UMPU) $\alpha$-level test procedure. In Neyman-Pearson (NP) hypothesis-testing theory, if this "optimal" (UMPU) function $R_A(\alpha)$ exists, nothing more is needed for the decision about $F_A$.

The maximum acceptable Type-I error α may be left unspecified, and the reader may be provided instead the observed decision P-value $p_{inf(α)} = \inf\{α: y ε R_A(α)\}$; their decision can then be based on inserting their own α in the P-rule "reject $F_A$ if $p_{inf(α)} ≤ α$" (Lehman, 1986, p. 70). This methodology is however usually implemented by finding a test statistic $T_A = t(Y;F_A)$ with realization $t = t(y;F_A)$ and mass function $h_A(t)$ such that $p_{inf(α)} = p_{t;A} = h_A(\{T_A ≥ t\}) = \int_{u≥t} h_A(u)du ≤ α$ if and only if $y ε R_A(α)$; in basic examples $T_A$ is a log likelihood ratio. Many sources however define a decision P-value for $F_A$ as the corresponding random variable $P_{T;A} = \int_{u≥T} h_A(u)du$ with realizations $p_{t;A}$. If $f ε F_A$ and $R_A(α)$ defines a valid test, $\Pr(P_{T;A} ≤ α) = f(\{Y ε R(α)\}) ≤ α$ and thus $P_{T;A}$ is distributed as or exceeds a unit-uniform random variable when $f ε F_A$.

The latter uniformity condition is sometimes used to define P-value validity, for it ensures that for any α the false-rejection rate (size) of the P-rule never exceeds α. Both the divergence $P_{D;A}$ and decision $P_{T;A}$ are valid in this sense; a key difference is that $P_{T;A}$ is further optimized for power, which can bring its distribution closer to uniform from above. Note however that there are no single-sample conditions applied to the realizations $t$ or $p_{t;A}$; this neglect is a source of defects of decision P-values as evidence measures, and may arguably be seen as a defect in their use as decision guides as well. An analogous objection to decision P-values arises when, due to discreteness, exact uniformity of P-values and an exact size α UMPU tests cannot be obtained without randomizing decisions at boundary observations. This leads to examples in which the single-sample decision comes down to a coin toss – an anomaly avoided by redefining the P-value (Lancaster, 1961).

*The two types of P-values are not equivalent*

Most of the literature seems to ignore that there are two definitions and interpretations of P-values, and to the extent it recognizes them, it treats them as differing only in that a divergence P-value is an observed (realized) quantity and a decision P-value is the corresponding random variable. This is not however true in general: When single-sample descriptive-geometric coherence criteria are imposed on discrepancy P-values (for example by using a divergence measure as the discrepancy) but only repeated-sampling criteria are used to derive decision P-values, we can have $p_{d;A} > p_{t;A}$, as in the example below where $p_{d;A} - p_{t;A}$ approaches ½. While such examples correspond to greater power for the P-rule, Schervish (1996) showed that, for $F_A$ defined by bounded parameter intervals and UMPU procedures, the resulting decision P-values could violate subset coherence (condition 2) in that one could have $F_M \subset F_A$ and yet have $p_{t;M} > p_{t;A}$.

In sum, realizations $t(y;F_A)$ of UMPU test statistics can violate basic axioms for divergence measures when treated as functions of $F_A$, leading to incoherent decision P-values $p_{t;A}$. This incoherency can be attributed to giving power priority over single-sample coherence, and arguably should disqualify $p_{t;A}$ as a measure of data evidence about A. In contrast, a divergence P-value $p_{d;A}$ starts from a coherent discrepancy measure and thus will coherently indicate the compatibility of the targeted constraint set A or family $F_A$ with the data y (more precisely, $f_y$), at least in large samples and often exactly, depending on the divergence measure. Consequently, $1-p_{d;A}$ or $-\log_c(p_{d;A})$ will coherently indicate the degree of refutation of A provided by the divergence measure – all without reference to a precise alternative.

Interestingly, Schervish took his result to imply that decision P-values were incoherent measures of evidence *supporting* $F_A$. As will be discussed below, there are purely logical reasons for rejecting any statistic as an absolute or unconditional measure of support. As likelihoodists often argue, support (and thus the possibility of acceptance of A) can only be relative to or conditional on specified alternatives to $F_A$ (e.g., Edwards, 1992; Royall, 1997). Similarly, decision procedures also need to specify alternatives to $F_A$, although this need does not dictate imposition of UMPU (Hansen & Rice, 2023).

*Summary*

Divergence P-values summarize probabilistic information about how far the data diverge from an embedding model A, or how far the data filtered through an embedding model A diverge from the data filtered through a more restrictive target model M. The information is measured using the geometry of the model family defined by A or M. The definitions of divergence P-values differ from NP definitions: In NP statistics, observed P-values are intermediate calculations for decisions derived solely to satisfy resampling criteria; they are defined as the minimum α-level at which the tested model would be rejected by an optimized NP test (Lehmann, 1986, p. 70). In this sense, claims that P-values are incoherent or do not measure evidence are circular, being consequences of adopting definitions that force P-values to conform to repeated-sampling criteria (such as UMPU) without regard to single-sample coherence. In contrast, definitions derived instead from geometric comparisons of observed data and its projections onto model regions will automatically satisfy at least approximate coherence requirements.

**Misconceptions about P-values and evidence**

There are now hundreds of articles spanning many decades about how everyday researchers misinterpret and abuse statistical tests and P-values. There are far fewer about how myths about P-values have been promulgated and perpetuated. For example, it is not uncommon to see claims that P-values "overstate evidence against null hypotheses". Yet observed P-values are just basic probability statements about the location of a statistic in a reference distribution, mute about their own "strength". Thus, the claims do not reflect anything intrinsic to the concept of a P-value, but instead reflect a synergism between user mistakes and philosophical commitments that argue against use of P-values: Users mistakenly perceive the limited evidence encoded in observing p=0.05 as strong simply because 0.05 is an entrenched standard for declaring "statistical significance" – and, far too often, a decisive factor for publication, as illustrated in Fig. 1 of van Zwet & Cator (2021). In tandem, some statistical commentators take posterior probabilities or likelihood ratios as "gold standards" for evidence measurement; those standards are however rejected by those who see how the conflict of those "standards" with frequentist testing can reflect failings of the prior or likelihood models, rather than deficiencies of the frequentist assessments (Ritov et al., 2014; Greenland, 2019a; Bickel, 2021a, 2021b).

A more subtle cognitive problem arose however as modern mathematical statistics emerged. Originally, P-values were defined directly as probabilities of observed events, where the events concerned statistics that gauged data divergence from expectations computed from a target (test) model or hypothesis (e.g., Pearson, 1900; Fisher, 1934, p. 66). These original P-values made no reference to alternative models or formal decision rules (Cox, 1977). That view further recognized how closeness of the model predictions to the data (as indicated by a large goodness-of-fit P-value) did not imply the target model is correct (Pearson, 1906). But Neyman-Pearson (NP) hypothesis testing altered the definition of a P-value to that of a type of random variable P whose realization p was the smallest α level at which the criterion "reject if p≤α" would reject the hypothesis, given the observed data (Lehmann, 1986, p. 70) – a redefinition which almost inextricably bound P-values to statistical decisions.

Many subsequent authors wrote as if the original goodness-of-fit and NP definitions were mathematically equivalent, even though (as will be discussed below) this not always the case. This subtle mistake parallels but is distinct from two different traditions. One tradition uses P-values as continuous evidence measures, and again is often called "neo-Fisherian" although traceable to Karl Pearson (Hurlbert & Lombardi, 2009). In contrast, the NP tradition uses P-values solely as components of decision criteria, a usage which requires specification of alternatives along with choice of a cut-off or α-level justified by error costs (Neyman 1977; Lakens et al. 2018, Mayo 2018) – albeit in practice a justification is rarely seen and instead a default α (most often 0.05) is used with no regard to actual error costs or to uncertainty about the model used to justify error-rate claims.

With the ascendance of NP theory in mathematical statistics, many if not most writers now ignore how P-values can be defined and treated descriptively, unconnected to formal decision rules; they instead focus on how P-values can malfunction when (as often the case) they are misused to measure support, or else focus on how they can be properly connected to Bayesian measures. The present paper shows how descriptive P-values can be derived to provide coherent measures of refutational evidence, without invoking notions of support.

*Descriptive P-values measure compatibility without measuring support*

The scale of a P-value p typically runs from 0 to 1, where 0 implies impossibility of an observed statistic under the target model (complete discord between the statistic and the model) and 1 corresponds to no conflict of the statistic with the model (no discord between the statistic and the model). This scaling seems to suggest P-values measure support, but again only reflects degrees of compatibility between the data and the model. The description of P-values as compatibility measures can be seen in Box (1980, p. 387), Bayarri & Berger (2000, 2004), Robins et al. (2000) and Greenland (2019a), and is anticipated by this passage in Fisher (1935, p. 207; emphasis added):

> "If we consider a series of hypotheses with different values for the diminishing return [on crop yield from adding more phosphate fertilizer], and determine which of these values are **compatible**, at any given level of significance, with the observed yields, some of the values which would appear to be acceptable would be negative…";

see also Pearson (1900, p. 170-171) and Fisher (1934, p. 66) for similar use of "compatible" with P-values. The same concept has also been described as goodness-of-fit (Pearson, 1900), consonance (Kempthorne & Folks, 1971; Folks, 1981), and consistency (Cox, 1977); unfortunately, "consistency" is more often used for unrelated convergence properties. Other terms for the compatibility concept include conformity, concordance, and accord (Rice, personal communication). Its opposite may thus be termed incompatibility or discordance, which can be measured by 1−p, although there are cognitive and abstract arguments for instead employing the surprisal measure or S-value $s = -\log_c(p)$ for incompatibility (Greenland, 2019a; Rafi & Greenland, 2020; Greenland 2021a; Amrhein & Greenland 2022; Greenland et al., 2022, 2023).

Understanding why compatibility is a logically weaker concept than support requires recognizing why conformity of observations to a target model – that is, failure of observations to conflict with or refute a model – does *not* by itself imply support for the model: There are always many other models that will fit the data just as well yet contradict the target model; consequently, one must *a priori* impose very tight restrictions on those other models to generate a support measure. As an extreme example, upon observing no difference between treated groups in a randomized trial, we may say the model of no treatment effect is perfectly compatible with the data; but unless we restrict alternative models to trials that are competently and honestly conducted, we must face that the same observation is also perfectly compatible with models in which the data were mishandled or altered to erase the appearance of an effect.

In going beyond the simple realm of binary logic and into a world of uncertain and extensive complexity, support and conflict are not logical complements of one another because they do not encompass all the possibilities for evidence. For example, evidence may be wholly indeterminate, supplying neither conflict with nor support for a targeted hypothesis. The conceptual incompleteness of conflict and support corresponds to the asymmetry between evidence against and evidence for a hypothesis, which is emphasized heavily in falsificationist philosophy (e.g., Popper, 1959); there, "corroboration" is used to indicate failure to refute a hypothesis without implying support or confirmation. Corroboration is meant to be weaker than support; nonetheless, in ordinary language, compatibility is even weaker than corroboration, for compatibility suggests nothing at all about the power or severity of the criterion being used for evaluation – it is simply the opposite or negation of incompatibility. For example, making one coin toss and observing heads is highly compatible with the hypothesis H: "The probability $\theta$ of heads is ½"; but to say it corroborates or supports H insinuates that the observation has more than trivial evidential value about that H. On the other hand, incompatibility and refutation do have a parallel: Most would say observing heads is highly incompatible with and even refutes H: "$0 < \theta \leq 10^{-9}$" even though heads is not impossible under H.

Ignoring the difference between compatibility and support can be seen as a source of claims that evidence measurement and statistical testing require specification of alternatives, one of the major points of contention in the Fisher-Neyman split (Fisher, 1955; Pearson, 1955; Neyman, 1956). While addressing this asymmetry is not needed for the mathematical development, it should be borne in mind when interpreting results. In particular, a reason to refer to P-values as compatibility measures is that they are more accurately understood when recognized as reversals of conflict orderings, rather than as measures of support (which require restriction of alternatives to a well-defined family of distributions).

*What do P-values and their transforms describe?*

P-values describe a relation of a model to data, or the relation of a more restrictive model M to a less restrictive embedding model A in light of the data, as seen in comparing observed to expected data, or more generally in comparing M-expected to A-expected fits to the data when M is nested in A (i.e., imposes all the constraints in A and more). The latter comparison measures information about M in the data, given A. Another way to characterize this process is that of measuring divergences among different degrees of data smoothing, ranging from no smoothing (the raw data, which is the expected data under a saturated model) to A-smoothing to more severe M-smoothing.

The greater the smoothing (the more constraints imposed), more data detail will be discarded as "noise" under the model, and more certainty will be assigned to any structure (apparent signal) remaining in the expectations (e.g., that a parameter appears to be large enough to signal treatment superiority, where the latter is defined context-specifically). Of course, the model may be mistaken, in which case important information may be discarded or smoothed away by the model, and excessive certainty will be assigned to the remaining apparent structure (signal). Thus, to avoid catastrophic loss while getting rid of obvious noise and hopelessly weak signals, one may maximize the embedding model (minimize the number of constraints in A) based on context-specific information (Greenland, 2006).

Given a model, inference may be based on a model diagnostic, as when judging whether too much structure is smoothed away (too much information is discarded) by the model using tools like residual plots and P-values for model fit. Upon adopting a working model that has passed these tests, inference becomes instead model conditional, as when model expectations are substituted for the data and then used to make claims about the data generator. Examples include marginal (population-standardized) effect estimates with components estimated from model-fitted quantities such as balancing scores. The inferential claims are often supported by P-values for coefficients or their inversion into interval estimates using the selected model, without accounting for the model selection that took place. As has long been documented, such preselection will invalidate decision rules that ignore it (Leamer, 1978). A pertinent question is then: Do these P-values mean anything despite the selection? One answer is that Fisherian P-values (those defined from observed divergences) still describe an observed discrepancy between the data and the model when that discrepancy is adjusted ("standardized") to allow for the random component of the model, which is to say the noise distribution assumed by the model. This interpretation holds even though the naïve comparison of such post-selection P-values to a pre-specified α would lead to repeated-sampling-and-selection error rates far higher than α, and thus brings to the fore the conflict between descriptive and decision uses of P-values.

I have avoided using "population" because it suggests to most users that the P-value is describing some target population from which the data were sampled. In reality, the data often come from a source that happens to be available but whose connection to the actual decision target may be quite speculative. Then too, the data are affected not only by intentional (interventional) design elements such as matching and randomization, but also by unintended influences such as causes of refusal to participate, loss to follow-up, and other procedural problems, as well as sample and data alterations made in the course of analysis (e.g., discarding units declared as outliers, or fixing records that have missing or clearly wrong data items). Thus, to validly infer back to the data source or make decisions with known error rates requires more than just rigid adherence to a pre-specified analysis protocol; it also requires a model which adequately reproduces the behavior of the entire physical process that determined the data used in the analysis (Greenland, 2005, 2022).

*Measuring compatibility and conflict with reference distributions and S-values*
In general form, the coherence criterion requires that a measure of evidence against a target model M or hypothesis H is never less than the measure against an embedding model A that contains the target model M as the special case of A in which the hypothesis H holds. A special case of this criterion was introduced above as subset coherence for P-values, which requires that the observed P-value for the fit of a target model M never exceeds the observed P-value for the fit of a more general, flexible embedding model A in which M is nested. An analogous criterion for a support measure requires that the measure for M never exceeds that for A. Such criteria can be traced at least back to Gabriel (1969) and have been extensively studied since, e.g., see Schervish (1996), Fossaluza et al. (2017), Bickel & Patriota (2019), and Hansen & Rice (2023).

A descriptive P-value provides the location of an observed statistic in a reference distribution derived from the targeted ("test" or "null") model M *and* the data, and is thus tailored to the study as observed.

In this sense, it is anchored in the single-sample viewpoint. The reference distribution need *not* be based on the goal, structure, or optimization of a decision over repeated sampling; instead, its purpose is to provide coherent measures of compatibility and conflict between the data and the model. It may be highly conditioned on the observed data in ways that depend on the targeted model; hence the reference distribution may vary considerably across resampling from the distribution entailed by the actual data-generating mechanism. Again, as per Cox (1977, sec. 2.1), "such a procedure is to be distinguished sharply from a decision problem in which 'acceptance' or 'rejection' is required".

In contrast, in Neyman-Pearson (NP) statistics, P-values are random variables that are components of the decision rule "reject if P≤α", or are by-products of the decision "reject if the test statistic falls in a critical region of size α" (Lehmann, 1986, p. 70). In the pure frequentist view of Neyman (1977), losses and hence these random variables are evaluated over their actual resampling distributions according to criteria such as power and "unbiasedness", without regard to features of single-sample realizations; they thus can suffer from incoherencies when extended to interval hypotheses (Schervish, 1996; Hansen & Rice, 2023). Statistical arguments against P-values as evidence measures have been based on these decision-theoretic definitions and criteria; they have largely ignored or dismissed treatments that start from the original divergence definition (Pearson, 1900) to interpret P-values or their transforms as single-sample information summaries or measures of compatibility or conflict between data and assumptions. The sections below will illustrate and expand on how the descriptive treatment of P-values differs from the decision-theoretic treatment, focusing on the special case in which a P-value summarizes relations among models and data within an information geometry defined by the observations as well as the models.

The Appendix provides a technically more detailed review of divergence P-values illustrated with basic generalized-linear models (GLMs). In doing so its focus is on intuitive visualization of actual single-sample properties of P-values. It further reinforces arguments that common misperceptions of P-values can be mitigated by transforming an observed p to an unbounded and reversed scale to provide a measure of incompatibility, conflict, refutation, or surprise. This rescaling also removes evidence measurement from the 0-to-1 scale on which posterior probabilities live, thus removing a source of confusion of P-values with posterior probabilities.

The most common example of such a transform is the negative base-c logarithmic transform to a surprisal or S-value $s_c = \log_c(1/p) = -\log_c(p) = -\ln(p)/\ln(c)$; $s_c = 0$ then says the observed discrepancy statistic or divergence measure from which the P-value was computed was completely unsurprising given the target model, or that it does not conflict with or contains no information refuting the model. As the divergence increases, $s_c$ increases to reflect surprise at the data given the model, or conflict with the target model (Bayarri & Berger, 1999; Greenland, 2019a; Rafi & Greenland, 2020; Cole et al. 2021; Gibson, 2021; Amrhein & Greenland, 2022; Greenland et al., 2022). The base of the log, c, can be seen as a scale factor; c=10 is commonly used but has no theoretical justification. Taking instead c=*e* produces the natural S-value $s_e = -\ln(p)$ which is a special case of the "E-value" in Grünwald et al. (2021) and the "betting score" in Shafer (2021); see Greenland (2021a). As explained in the Appendix and elsewhere, c=2 provides an intuitive mechanical interpretation: The binary S-value $s_2 = -\log_2(p)$ measures bits of information against a model supplied by the statistic, which translates easily into outcomes of coin-tossing experiments (Greenland, 2019a; Rafi & Greenland, 2020; Cole et al. 2021; Greenland 2021a; Greenland et al., 2022).

**Competing concepts of P-values**

*Coherence vs. decision P-values*

For NP frequentists, statistical decisions calibrated to long-run frequencies are the central goal of analysis, and the paramount definitions and evaluation criteria refer only to resampling frequencies. Pursuing this goal, modern NP theory treats a P-value for a hypothesis H as a random variable that follows (or at least approximates from above) a uniform resampling distribution when H is correct, and concentrating toward 0 otherwise. This theory makes no reference to data description; in practice however the ensuing single-sample decisions are usually misrepresented as data descriptions, as when a paper reports "no association was observed" when in fact this only meant that p>0.05.

An acceptable measure of incompatibility or conflict between data and hypotheses must obey the single-sample criterion of subset coherence, in that its value for a hypothesis expressed as a subset of a function or parameter space cannot exceed its value for any of its own subsets. In parallel, a measure of compatibility is *coherent* if its value for such a hypothesis cannot exceed its value for any of its supersets. Again, compatibility is a weaker condition than support, in that support implies compatibility but compatibility does not imply support. Compatibility and support do however share an analogous coherence concept: A measure of support by data for a hypotheses is *coherent* if its value for a hypothesis cannot exceed its value for any of its supersets. Likelihood maxima over sets satisfy this requirement and thus are coherent support measures.

In contrast, using interval hypotheses in a simple normal location model, Schervish (1996) showed that an extension of uniformly most powerful unbiased (UMPU) P-values to interval hypotheses (Lehmann, 1986, sec. 4.2) is an incoherent measure of support; hence a scale reversal such as 1−p or −$\log_c(p)$ would provide an incoherent measure of conflict if p were defined from an UMPU test of an interval hypothesis. Similar problems can arise with Bayes factors (Lavine and Schervish, 1999), but see also Good (2001). The UMPU extension goes beyond ensuring that the simple decision rule "reject if p≤α" has Type-I error rate (size) no more than α over the resampling distribution under the interval hypothesis, by imposing conditions to optimize power. Key to the present discussion is that the P-value in the rule for an interval UMPU test (Hodges and Lehmann, 1954; Lehmann, 1986, sec. 4.2; Schervish, 1996) is defined and evaluated solely over resampling, with no attention to single-sample coherence.

*Divergence P-values*

Incoherence does not afflict all extensions of P-values, such as those that aim only to measure information on a compatibility/incompatibility scale. Consider definitions that lead to maximizing a point-hypothesis P-value over a hypothesized interval for a real-valued parameter: Such max-p (or supremum) extensions have been disqualified from being "frequentist" by adherents of the NP definition (e.g., see Remark 1 of Robins et al., 2000 p. 1145). One reason is that the maximum two-sided P-value will equal 1 whenever the parameter estimate falls in the hypothesized interval, resulting in a distribution far above uniform when the parameter is in the interval interior. This property led Schervish (1996, sec. 4) to described max-p extensions as "simple minded" and "not particularly useful". But these dismissals are based on criteria that can be rejected by those who use P-values as discrepancy measures or information summaries rather than as test criteria. As illustrated below and explained in the Appendix, summarization can start from a geometric derivation of P-values from divergence measures. Doing so naturally leads to maximizing P-values over intervals, thus enforcing their coherence; for point hypotheses such as H: μ=m, these P-values simplify to ordinary two-sided P-values.

The descriptive goal is to summarize data information about models, particularly comparisons of fitted models against data and each other. A descriptive P-value is thus the *observed* quantile at which a discrepancy statistic sits in a reference distribution derived from the model. In particular, when divergence statistics are used to measure discrepancies, the reference distribution may be derived from a geometry on a model space that combines information on both the data-generating distribution and the observations. This distribution serves only to locate the data relative to the model in a particular direction in the model space, using a coordinate scaling standardized to the estimated sampling variation. The resulting P-values are numerically identical to observed decision-theory P-values when the divergence statistic and reference distribution chosen for summarization are identical to the test statistic and resampling distribution chosen for decisions. While this identity is most common, it is not dictated by the goals of the two approaches and is violated in the examples below and in the Appendix; Bayarri and Berger (2004) discuss conditioning as a method of reconciling the approaches.

As illustrated in the Appendix, the descriptive goal can be rephrased as that of depicting relations among different degrees of data smoothing or filtration, where the filtrations are through nested models. This goal stands in sharp contrast to Bayesian methods, whose goal is to make statements (or "inferences") about parameters conditional on data; these methods require inputs of fully specified parameter distributions and can break down spectacularly in high dimensions (Robins and Ritov 1997; Ritov et al. 2014). In terms of goals, NP decision theory is more akin to Bayesian methods in going beyond descriptions to make direct statements or take actions about parameters or models "in light of the data", although they differ from ordinary Bayesian decisions in allowing statements or actions that are not based on parameter distributions or full conditioning on the observed data.

A core requirement of NP decision theory and frequentist inference procedures is that the data generator involves a randomizer (for selection into the data, if population inference is a goal; or for treatment assignment, if causal inference is a goal) that is known apart from an estimable parameter vector (such as the coefficients in a sample-selection or treatment-assignment probability function). In the descriptive view, however, the randomizer is demoted to being simply another assumption in the model used to compute the P-value, and thus (as with other assumptions) its violation may be brought forward as an explanation for the size of an observed p. There is then no inference beyond the noting that a small p can signal a problem with one of the assumptions in the model used to derive the reference distribution – it may be small because the treatment has an effect different from that assumed by the model, or because there were uncontrolled nonrandom elements in the actual treatment assignment, observation selection, or data missingness (e.g., nonrandom censoring). This unconditional descriptive view does not *logically* condition on any assumption, and thus is inferentially mute: It does not single out violation of an assumed (test) hypothesis to explain why p seemed small; nor does it single out satisfaction of the hypothesis to explain why p seemed large. A descriptive P-value simply states a factual relation between what was observed and what was expected based on the target model, where "target model" refers to all the assumptions or constraints used to compute the P-value, including H.

The main point is that the conflicting goals of decision and descriptive frequentist approaches lead to definitions and interpretations of extended P-values that differ and may come into conflict. The two views can however be at least partially reconciled through the criteria and preferences they share. Consider the following passage from p. 1144 of Robins et al. (2000; emphasis added):

> …a p value is useful for assessing compatibility of the null model with the data only if its distribution under the null model is known to the analyst; otherwise, the analyst has no way of assessing whether or not observing p = .25, say, is surprising, were the null model true. **That we specify that distribution to be uniform is largely a matter of convention**.

The view presented here agrees in that, in order to properly interpret a P-value, we must know sufficient detail about the meaning and distribution of its source statistic under the target model and its violations. Nonetheless, most of the assumptions that compose the target model cannot be assured to hold in observational studies or in randomized experiments with much drop-out or nonadherence. Furthermore, when the target model does not involve dimension reduction (so the model has interior points in the information topology described in the Appendix), uniformity of a random P-value becomes a useful property only at boundary points of the distribution subspace defined by the targeted model.

For evaluating statistical performance, the descriptive goal further replaces power by information content. The descriptive view thus regards the emphasis on uniformity of P-values under assumed models (as seen in most theoretical literature) as a product of NP goals applied to highly idealized experimental models. Because those models are usually unrealistic in health and medical research, we seek instead to describe the relations that hold between observations and models, regardless of the model's accuracy.

*Example: A divergence P-value for a simple interval hypothesis*
Consider the model for n observations arrayed in a vector $y = (y_1,...,y_n)'$ in which the $y_i$ are assumed to be independent draws from a normal distribution with unknown mean $\mu$ and known variance $\sigma^2$; call this set of assumptions the embedding model A, leaving µ the only free parameter in the model. This embedding or background model places severe restrictions on the distribution F(y) for the random data vector Y = $(Y_1,…,Y_n)'$: It says that if the physical data generator were left to run indefinitely, it would behave exactly like a random-vector generator for an uncorrelated n-variate normal distribution with equal $Y_i$ means and variances. Note that A implies the sample mean of the $y_i$ can replace the data without loss of information about the data generator, i.e., it is sufficient for determining the behavior of the generator to the extent allowed by the data, since A implies that behavior is normal (Gaussian) with variance $\sigma^2$. But A does not imply that the actual observed-y histogram would be visually well approximated by a normal density with mean equal to the sample mean and variance equal to $\sigma^2$, which is often what is meant when saying the fitted model is adequate as a data summary or compression; the latter claim requires evaluation of A against the full data vector y, including graphical as well as goodness-of-fit diagnostics.

Accepting A for the moment, we may wish to evaluate the information that the data vector y supplies against a submodel M nested within A that imposes additional restrictions H on µ. Specifically, suppose M adds to A the restriction H: µ=m with m known. Evidence against µ=m is usually gauged by finding the "standardized" distance $|\hat{\mu}-m|/(\sigma/n^{½})$ from the sample mean $\hat{\mu}$ to the target value m in a standard-normal tail-area function to obtain the 2-sided P-value $p_m$. Equivalently, to get $p_m$ we could find the divergence statistic $d_m = d(m;\hat{\mu}) = (n/\sigma^2)|\hat{\mu}-m|^2$ in a 1 degree-of-freedom (df) $\chi^2$ distribution F(d), whence we can describe the standardization as multiplying the squared distance from the data summary to the hypothesized mean m by $n/\sigma^2$, which is the amount of Fisher information in the data about µ given the model A. Furthermore, given normality, $d(m;\hat{\mu})$ is identical to the usual likelihood-ratio (LR) statistic for evaluating M against A, and (as discussed in the Appendix) is twice the Kullback-Leibler information divergence of M from A.

Under M, the distributions of the random analogs $D_m$ of $d_m$ and $P_m$ of $p_m$ are derived by treating the $Y_i$ as independent normal($m,\sigma^2$); $P_m$ will then be uniform, implying the NP decision rule "reject H if $p_m \leq \alpha$" will have size α; i.e., Pr($P_m \leq \alpha$;µ=m) = α. If however µ≠m but A (i.i.d. normality with known variance) still holds, the distribution of $P_m$ will be increasingly concentrated toward 0 as the noncentrality parameter $d(m;\mu) = (n/\sigma^2)|\mu-m|^2$ grows.

The random divergence statistic $D_m$ and P-value $P_m$ for H: $\mu=m$ are identical to the test (decision) statistic and P-value in NP theory, and their distributions are fixed and known given $\mu=m$. The divergence view departs from the testing (decision) view when M instead adds to A an interval restriction H: $m_L \leq \mu \leq m_U$. The squared distance from $\hat{\mu}$ to $[m_L,m_U]$ is

$$d([m_L,m_U];\hat{\mu}) = \min\{d(m;\hat{\mu}): m \in [m_L,m_U]\}$$

The resulting observed divergence P-value for the model M or for the hypothesis H given A is

$$p_M = \max\{p_m: m \in [m_L,m_U]\}.$$

When $m_L = m_U = m$, we get $p_M = p_m$. But when $m_L < m_U$, the distributions of $d([m_L,m_U];\hat{\mu})$ and of $p_M$ are no longer known given M; they can only be computed conditional on $\mu$ and may vary considerably across $\mu$ in the interval $[m_L,m_U]$ defined by H.

We can nonetheless summarize the discrepancy of $\hat{\mu}$ from $[m_L,m_U]$ using $p_M$. The divergence $d([m_L,m_U];\hat{\mu})$ goes to infinity as $\hat{\mu}$ becomes more distant from $[m_L,m_U]$, and is at its minimum of zero when $\hat{\mu}$ is in $[m_L,m_U]$. Thus, $p_M$ will range from zero ($\hat{\mu}$ *infinitely far* from H) to one ($\hat{\mu}$ *zero distance* from H). If we consider the standardized distance $d(m;\hat{\mu})$ of $\hat{\mu}$ from the interval as indicating the extent of incompatibility between the observations and the hypothesis, $p_M$ can be taken as an index of compatibility of the data summary $\hat{\mu}$ with the model M that restricts $\mu$ to $[m_L,m_U]$. To restore a direct relation to the divergence and incompatibility, we may transform $p_M$ to a surprisal or S-value such as $s_M = -\log_2(p_M)$, which ranges from zero: $\hat{\mu}$ *zero distance from the interval* to infinity: $\hat{\mu}$ *infinitely far*; $s_M$ provides other benefits for interpreting the observed divergence in terms of the information it supplies against H given A (Greenland, 2019a, 2021a; Rafi & Greenland, 2020; Cole et al. 2021; Greenland et al., 2022, 2023).

Turning to the corresponding random P-value $P_M$, we see that with correct A, $m_L < m_U$, and increasing sample size

- when H is incorrect, $\mu$ is exterior to the interval ($\mu<m_L$ or $m_U<\mu$) and $P_M$ converges to 0;
- when $\mu$ is interior to the interval ($m_L<\mu<m_U$), $P_M$ converges to 1;
- when $\mu$ is a boundary point ($m_L$ or $m_U$), $P_M$ converges to a 50:50 mixture of a uniform(0,1) variable and a point mass at 1.

From the last item we see that $P_M$ behaves quite differently from an NP-optimized P-value; this is so even if we adjust $P_M$ to make it approximately uniform at the boundaries. For example, if $m_L < m_U$ and H is incorrect, the probability of seeing a positive divergence on the wrong side of the interval will become negligible as n increases; hence the use of the two-sided $p_m$ to define $P_M$ reduces the asymptotic power of a test that treats $P_M$ as if uniform under H. But if H is correct and $\alpha<½$, the size of the rule "Reject H if $p_M \leq 2\alpha$" will converge to $\alpha$ at the boundary points, and for half intervals ($m_L = -\infty$ or $m_U = \infty$) it will be exactly $\alpha$ at the finite boundary – although in both cases, at interior points the rule will become increasingly conservative as n grows, with size going to zero. Extensions of $p_M$ and generalizations of these results are illustrated in the Appendix.

*Comparison of the divergence and UMPU P-values*

Continuing the example, **for $\hat{\mu}$ exterior to the open interval ($m_L,m_U$)** the Hodges-Lehmann (1954) UMPU decision P-value $p_{HL}$ for H in Schervish (1996, eq. 2) can be written as the mean of the two-sided P-values for the boundary points: $p_{HL} = (F_1[d(m_L;\hat{\mu})] + F_1[d(m_U;\hat{\mu})])/2$ where $F_1$ is the 1 df $\chi^2$ cumulative distribution (not upper tail). **This formula does not give $p_{HL}$ for interior points; instead, it shows that, for $\hat{\mu}$ outside the interval, $p_{HL}$ is the mean of the divergence P-values for the interval endpoints $m_L$ and $m_U$. This is in contrast to $p_M$, which for $\hat{\mu}$ outside the interval is the maximum of the divergence *P*-values for those endpoints.**

When $m_L = m_U = m$, both $p_{HL}$ and $p_M$ equal the two-sided P-value $p_m$ for $\mu=m$; but with $\hat{\mu} = m_U$, as $m_L$ approaches $-\infty$ $p_{HL}$ approaches the one-sided P-value for $\mu = m_U$ while $p_M$ approaches the two-sided P-value for $\mu=m_U$ and thus approaches $2p_{HL}$. Suppose now the interval H is $[-\sigma,\sigma]$ and consider some numeric properties of these observed statistics in single samples of size n. With n=1 and $\hat{\mu} = \sigma$, $p_{HL}$ would be 0.523, approaching 0.50 as n increases; while for fixed n, $p_{HL}$ would achieve a sharp maximum when $\hat{\mu} = 0$. This is in stark contrast to $p_M$, which is at its maximum of 1 for any $\hat{\mu}$ in $[-\sigma,\sigma]$. When $\hat{\mu}$ is in the interval interior, the difference decreases as n grows, in that $p_{HL}$ approaches 1 and thus approaches $p_M$; nonetheless, $p_{HL}$ instead approaches $p_M/2$ when $\hat{\mu}$ is near a boundary or exterior to the interval.

The message of such single-sample numeric illustrations is not that one of these P-values is correct and the other incorrect, but rather that they address fundamentally different goals and thus should not be expected or forced to agree in all cases. In particular, in decision theory the indicator of $p_{HL} \leq \alpha$ is used as surrogate for the unobserved indicator of H. In classical frequentist theory, the H indicator is either 1 or 0 according to whether H is correct or not, and thus is not a function of the observations at all; the only concern is with the error rate of the rule used to impute this indicator over the entire space of possible samples. In contrast, $p_M$ can be used as a direct description of a single sample, expressing where the observed μ̂ fell relative to the model region defined by H, with $p_M = 1$ when μ̂ is inside the region but tapering off toward 0 as μ̂ moves away from the region based on the distance-scale factor $\sigma/n^{\frac{1}{2}}$.

In the above interval examples and more general ones discussed in the Appendix, the ratio between the divergence and UMPU P-values is bounded above by 2. A two-fold P-value ratio would translate to power loss if $p_M$ were substituted into the decision rule "reject if $p\leq\alpha$", as produced by observed $p_{HL}$ and $p_M$ on opposite sides of α. For example, 80% power for a 0.05-level UMPU test could be reduced to 71% if $p_{HL}$ were replaced by $p_M$. But descriptively the ratio of $p_M$ and $p_{HL}$ represents at most a difference of $\log_2(p_{HL}/p_M) = 1$ bit of Shannon information against H. Thus, the apparent single-sample information gain from imposing UMPU is equivalent to the information in one coin toss against fairness of the tossing set-up, and is purchased by sacrificing single-sample coherence.

Another case in which imposing UMPU leads to conflict with descriptively sensible divergence P-values arises in equivalence testing, where two-one-sided test (TOST) procedures are often used despite being biased in finite samples. TOST procedures can even have zero power (Berger and Hsu, 1996 sec. 4.2). Consider again the normal-mean example with known σ, n=1, and the nonequivalence hypothesis H: $\mu\leq-\sigma$ or $\mu\geq\sigma$. The alternatives to H are now the μ in the open equivalence interval (−σ,σ) of radius σ. The TOST procedure uses the ordinary one-sided P-values $p_L$ for $\mu\leq-\sigma$ and $p_U$ for $\mu\geq\sigma$ in $p_{TOST} = \max(p_L,p_U)$ to produce the rule "reject H if $p_{TOST}\leq\alpha$"; since $p_L$ and $p_U$ are half the 2-sided P-values for μ=−σ and μ=σ, this rule is the same as "reject H if $p_M\leq 2\alpha$". Because $p_{TOST}$ and $p_M$ have minima of 0.16 and 0.32 at μ̂ = 0, the procedure cannot reject H (has zero power) if α **<** 0.16. Such zero-power examples properly reflect how the equivalence interval is too narrow to allow the sample mean μ̂ to diverge from H by more than the interval radius, which is scaled in units of $\sigma/n^{\frac{1}{2}}$. (As an aside, the term "equivalence test" can be somewhat misleading insofar as H is a hypothesis of *nonequivalence*; the test is thus most accurately described as a test *of* nonequivalence, or as a test *for* equivalence.)

**Discussion**

*Divergence Statistics for Descriptive Model Geometry*

To repeat, the goal of descriptive statistics is not to optimize decision rules, but rather to summarize discrepancies between model implications and observed data, or between data fitted under nested vs. embedding models (sometimes expressed as comparing data under different degrees of filtration). Thus, while descriptive P-values are derived from sampling distributions for discrepancies, their purpose is to summarize and transmit data information about relations among different models (which involve different distributions) in terms of the random variation allowed by the models. Decisions about the models require further methods and inputs such as loss functions. For related views see Cox (1977), Box (1980), Bayarri & Berger (2004), Bickel & Patriota (2019), Peskun (2020), Gibson (2021), and Vos & Holbert (2022).

There may be many choices for a discrepancy measure; the use of a divergence statistic is a natural one for descriptions based on visual geometric analogies and information measurement. The Appendix outlines technical details of divergence-information P-values and interval estimates for common regression models, and then discusses their consequences for statistical conceptualization and practice.

*Relations to Bayesian and likelihood methods*
The present development is explicitly non-Bayesian as well as non-decision-theoretic; description is its goal, not inference. One may present measures of model goodness-of-fit (or lack thereof) without specifying whether the fit is acceptable or not for some purpose, or whether the model is plausible, likely, or probable given contextual background information. Descriptive P-values can however be employed as empirical checks of Bayesian models by treating priors as random-effects distributions in a prior-predictive format (Box, 1980; Bayarri & Berger, 1999).

Many relations of observed P-values to posterior probabilities and Bayes factors have been delineated, usually in the form of numerical bounding relationships or calibrations under certain assumptions (e.g., Berger & Delampady, 1987; Casella & Berger, 1987ab; Sellke et al. 2001; Berger, 2003; Bayarri & Berger, 2004; Greenland & Poole, 2013ab; Held & Ott, 2016, 2018; Bickel, 2021a, 2021b; Fay et al., 2022). This literature has focused on simple one-sided and point hypotheses. For example, under mild assumptions the usual two-sided point-hypothesis P-value and hence $s_e = -\ln(p)$ translate into a lower bound for Bayes factors $-e \cdot p \cdot \ln(p) = s_e/\exp(s_e-1)$ (Bayarri & Berger, 1999; Sellke et al., 2001; Greenland and Rafi, 2019).

Of key relevance to the present discussion is the logical and structural asymmetry between the refutational evidence or information in frequentist P-values and the confirmatory evidence in relative likelihoods and posterior probabilities. This asymmetry arguably reflects a deeper divide between statistical methodologies than that between frequentists and Bayesians: Neyman-Pearson (NP) methodology requires consideration of power and hence (as with pure-likelihood and Bayesian methods) requires precise specification of data probabilities under alternative hypotheses. In contrast, precise alternatives are not essential to either the derivation or interpretation of divergence P-values; the same is true of the e-values developed by Peskun (2020), which again (along with divergence P-values) I would call a type of evidential P-value.

*Dealing with the discrepancy between received theory and common practice*
Most statistical primers present a purely descriptive definition of an observed P-value as the tail probability of an observed discrepancy measure given a target hypothesis or model, while advanced theoretical works use an NP definition as the smallest α at which that target would be rejected. The problem remains that too few sources note the difference between them in definition, purpose and criteria, let alone how the two can differ numerically when the target has interior points. Again, NP testing criteria are not needed to derive or justify divergence P-values, which were developed as a type of single-sample information or evidence summary, not as a NP repeated-sampling decision criterion. In particular, when a divergence P-value is not NP optimal, it can be superior for indicating features of the observed statistics. For the latter purpose, repeated-sampling criteria such as UMPU can be sensibly rejected when the authors wish to discourage decisions based on their study and limit analyses to describing relations of observations to models while accounting for statistical uncertainties. This attitude is often justified when formal statistics fail to capture crucial study complexities and uncertainty sources (Amrhein et al., 2019; Greenland, 2019a; McShane et al., 2019; Rafi & Greenland, 2020; Amrhein & Greenland 2022; Greenland et al., 2022, 2023).

A sensible real-world decision procedure requires more than a P-value: External inputs are needed, such as error tolerances or expected loss over an error distribution. Also essential is information on which (if any) assumptions used to compute the P-value can be safely taken as certain. Such inputs may require considerable effort to generate, and are usually unavailable to those reporting research results. This reality explains the need for the divergence view, which allows decisions and even inferences to be omitted when reporting statistical summaries such as P-values and interval estimates. In medical-research contexts some argue further that statistical tests *should* be omitted because few if any contextually sensible decisions can be made solely on the basis of statistical tests.

The position adopted here is that descriptive treatments are essential, even if they are followed by statistical decisions. To illustrate, consider that a P-value of $p_M$ for a model M in a study is analogous to a score at the $100p_M$ percentile of a college admissions exam when the score is placed in (standardized to) the histogram (reference distribution) of the scores of all applicants. An admission decision requires many considerations beyond this percentile, such as interview results, recommendation letters, and the number of positions available to set a rejection cutoff. The exam percentile is merely an ordinal measure of the examinee's performance on one exam, without further specifics such as whether it meets an admission requirement or how well it predicts the examinee's future performance. Likewise, the observed $p_M$ for M is an ordinal measure of the performance (fit) of the model in one trial, without further specifics to guide the use of $p_M$ for decisions; nor would it tell us how well the model would perform in subsequent uses.

*The overconfidence of confidence intervals*

The preceding observations have profound consequences for interval estimates computed from P-values by varying a target parameter in a model (typically a regression coefficient). Such intervals are routinely presented as "confidence intervals" with inadequate attention to how the claimed coverage rate requires certainty about various background assumptions used to compute the interval (Greenland, 2005, 2019b; Rafi & Greenland, 2020). The narrowness of such an interval is often misinterpreted as providing confidence that the parameter has been accurately estimated. Nonetheless, that narrowness may only reflect how faulty background assumptions produced the small P-values attached to parameter values outside the interval. The same caution against overinterpreting narrowness applies to Bayesian posterior intervals. To mitigate the harms of overconfidence in interval estimates, we have encouraged the renaming of such interval estimates to *compatibility intervals*, recognizing that without stringent and often unmet assumptions, the intervals only indicate which parameter values are compatible with the assumed background model (Greenland, 2019a, 2019b; Rafi & Greenland, 2020; Cole et al., 2021; Amrhein & Greenland 2022; Greenland et al. 2022, 2023).

*What is an accurate term for H?*

H is traditionally called a "null hypothesis"; but, following Neyman (1977), the present treatment rejects this label because it creates misconceptions among researchers that statistical hypotheses always correspond to there being no association or effect. This misimpression apparently arose because the ordinary English definition of "null" is zero or nothing, and so "null" is best avoided unless the hypothesis of interest really does correspond to no association or effect (Greenland, 2019a). In reality, H may imply the association or effect exists or has a particular nonzero value. For this reason, Neyman (1977) used the term "tested hypothesis" in place of "null hypothesis". Nonetheless, that term is inaccurate when measurement rather than testing is the analysis objective. A more accurate general term for H is *target hypothesis*, as adopted here.

Detailed arguments for replacing the traditional terms "significance level" by "P-value" or "α-level" as appropriate and for replacing "confidence" by "compatibility" have been given elsewhere (Amrhein et al. 2019; Greenland, 2019a; Rafi & Greenland, 2020; Amrhein & Greenland, 2022; Greenland et al., 2022).

*Related work*

Defenses of P-values as evidence measures along other lines continue to appear, e.g., Huisman (2022). The present development parallels that in Peskun (2020) in several ways, differing in that Peskun starts from tail areas rather than divergence measures to create a coherent unit-interval measure he calls an "e-value" ("e" for evidence). Unfortunately, the term "E-value" is already in widespread use in the medical literature for an entirely unrelated confounding bound (Sjölander & Greenland, 2022). In the present view, Peskun's e-value is a coherent evidential P-value that will very often equal the divergence P-value described here. Bickel and Patriota (2019) instead base development on a renormalized P-value they call a "c-value" ("c" for compatibility), whereas Grünwald et al. (2021) and Grünwald (2022) employ P-values that invert their notion of "E-value" (again, "E" for evidence).

Patriota (2013) presented a coherent unit-interval scaled measure of support for H he called an "s-value" which should not be confused with the surprisal S-value $s = -\log_c(p)$ described above. His development parallels that of the divergence P-value, which it often equals, and can depart from the NP P-value; but it is still decision-based in that it is found by minimizing confidence regions rather than divergence measures. Others have proposed coherent modifications of hypothesis tests derived from consideration of Bayesian decision theory (e.g., Fossaluza et al., 2017; Hansen & Rice, 2023).

**Summary and Conclusion**

It has been argued that P-values should not be used as measures of evidence, in part because some P-values violate coherence criteria in certain examples. These examples are however based on decision-theoretic definitions of and criteria for P-values, which overlook definitions and criteria based instead on summarizing geometric fit of observations to models. In geometric definitions, P-values are derived as descriptions of the divergence of a target model from data, or from the projection of data onto the embedding model. Their calibration then becomes a secondary process rather than a defining criterion. In this approach, P-values can coherently indicate the compatibility of the target model with the data or the embedding model, and can be transformed into coherent measures of incompatibility or refutational evidence. In simple cases, divergence P-values coincide with decision-theoretic P-values, but differ when the latter become incoherent. More generally, they avoid single-sample anomalies by placing descriptive criteria above the repeated-sampling criteria that lead to the anomalies, and can be used to reinterpret interval estimates as compatibility summaries rather than unjustified expressions of confidence.

As discussed in the Appendix, some reconciliation between the two types of P-values is possible by choosing P-values that satisfy criteria of both types. But again, coherence and UMPU criteria cannot be jointly satisfied in all settings, in which case arguably coherence should prevail. Regardless of the choice, it should be borne in mind that support is not the logical complement of refutation: Unlike support, refutation of a target model M does not logically require any alternative, because the negation "not M" specifies nothing other than that M is deleted from the vast realm of logical possibilities. And, unlike some decision P-values, divergence P-values can be used to measure refutational information (conflict) by taking their Shannon (negative log) transform or S-value as a measure of the minimum information against M provided by the observed divergence between M and the data or M and the the background model. Thus, while different theories can agree that frequentist P-values should not be used as measures of support, divergence P-values can be transformed into measures of conflict of refutation of hypotheses by data given background assumptions data with models or refutation of models by data. Divergence P-values can thus indeed coherently encode evidence *against* hypotheses or models.

**Acknowledgements**

I am grateful to Valentin Amrhein, Kenneth Rice, and Paul Vos for providing insightful comments and discussion on earlier drafts of this paper.

Corresponding author's address: lesdomes@ucla.edu (no physical mail address).

**APPENDIX: GEOMETRY OF DIVERGENCE STATISTICS IN REGRESSION MODELS**

**Notation and Concepts**

Following Fisher's original conception of observation space, a data set of n observations will be represented by a vector y in $R^n$ and treated as a realization of a random n-vector Y with an unknown distribution F(y) on $R^n$. Note that y standing alone represents the observed data, but y inside F(y) is an unspecified functional argument (dummy variable). Define a model for F(y) as a set A of assumptions expressed as constraints on F(y), including the distributional form of random components of Y and perhaps further restrictions such as equalities or inequalities among moments or model parameters. The model typically includes parametric constraints imposed by fixed covariates, for example via a design matrix and link function (McCullagh & Nelder, 1989). To minimize mathematical intricacies, it will be illustrated with the basic case of models indexed by their means, which include generalized-linear models (GLMs) for Gaussian regression with known covariance matrix, as well as ordinary binomial, multinomial, exponential, and Poisson regressions. As with common GLMs, I will assume throughout that A includes assumptions that render all elements of µ = E(Y) and cov(Y) finite with cov(Y) nonsingular; that µ fully specifies F(y) in that F(y;µ) is a known 1-1 function of µ, with F(y;µ) identified given A; that A does not bound the support of F(y) beyond the logical bounds on y; and that F(y;µ) has sufficient smoothness in y and µ for all statements to hold at least as asymptotic approximations.

Taking µ = y we obtain a distribution F(y;y) that reproduces the observed data vector y as its fitted mean vector; note that y appears in F(y;y) first as a dummy variable and second as the data. If A imposes no constraint on µ then A will be called a (mean-) saturated model; for all y, such models yield y as a maximum-likelihood estimate (MLE) of µ (other models may do so for particular but not most y). For simplicity I will assume throughout that F(y;y) is nondegenerate, as would be expected for count data with large samples, fixed n, and nondegenerate F(y;µ) (no structural zero counts, as per Bishop et al., 1975); for Poisson data this assumption translates to all $y_i > 0$. A more general approach which allows for random zero counts replaces y by counts $\mu_S$ fitted under a highly parameterized, minimally constraining model that removes zeros without oversmoothing the data, and then uses $F(y;\mu_S)$ instead of $F(y;y)$ (Greenland, 2006).

Let $R_\mu$ be the set of logically possible μ under the distributional form for $F(y;\mu)$ specified by A. Each point μ in $R_\mu$ then corresponds to a distribution $F(y;\mu)$ for Y in a distribution space $\boldsymbol{F}_\mu = \{F(y;\mu): \mu \in R_\mu\}$. I will assume the dimension $\dim(R_\mu)$ of $R_\mu$ is n; A may however include assumptions that logically constrain $R_\mu$. As examples, if A assumes that the Y components are independent normal then $R_\mu = R^n$; if A instead assumes that Y components are independent exponential or Poisson then $R_\mu$ is the positive orthant of $R^n$; and if A instead assumes Y components are independent Bernoulli then $R_\mu$ is the positive open unit cube in $R^n$.

Denote by $R_A$ the set of μ such that $F(y;\mu)$ satisfies all the constraints in A (including for example linearity constraints as well as logical constraints); $\dim(R_A)$ is often called the total degrees of freedom (df) for A and $\dim(R_\mu)-\dim(R_A)$ the residual df for evaluating A. As an example, suppose A says Y is multivariate normal(βx,Σ) with β an unknown scalar β, Σ a known positive-definite covariance matrix, and x an n-vector of known constants. Then $R_A$ is the line in $R^n$ through the origin traced out by letting β vary, $\dim(R_A)=1$, and the residual df is n−1. In contrast, for saturated A there is no constraint beyond the distributional ones, which implies $R_A = R_\mu$ and hence $\dim(R_A) = n$, leaving zero residual df for evaluating A.

**Nested model evaluations**

Suppose we wish to compare a more restricted model (larger set of constraints) M against a less restricted reference model A with fewer constraints, so that in terms of constraints $A \subset M$. In terms of the means μ however this containment is reversed to $R_M \subset R_A$, so that M is often said to be nested in or a submodel of A, and A is sometimes called the embedding or reference model for evaluating M. Classical goodness-of-fit evaluations of M against A such as Pearson's $\chi^2$ test take A as saturated with $\dim(R_A) = n > \dim(R_M)$ and define $df_{M|A} = \dim(R_A)-\dim(R_M)$ as the df for the evaluation . Nonetheless, as with interval hypothesis examples, we may have $df_{M|A} = 0$, yet evaluation is still possible.

The set of constraints added by M above A is the set difference H = A−M (usually $H_0$ is used, but that invites confusion with more specific constraints assigning zero to a parameter). M then equals the set union H+A, while A = M−H. A is sometimes called the set of auxiliary constraints for evaluating H. I will further assume that M confines μ to a fixed finite-dimensional subset of $R^n$ as n increases, and unless stated otherwise (as with saturated models) I will assume the same for A. The total df for H given A is sometimes defined as $df_{M|A}$ but again, this can be zero even if H is nonempty, as when H consists of inequalities.

As an example, suppose again the embedding model A includes (along with distributional assumptions) the constraint that μ = βx for an unknown scalar β and a known vector x of constants, so that $R_A$ is a line through the origin and $\dim(R_M) = 1$, while M adds above A the further assumption that β=0, corresponding to H: β=0 (equivalent to the constraint that μ is the zero vector). $R_M$ then is a single point (the origin) and $\dim(R_M) = 0$, while the df for H given A is $df_{M|A} = 1$. Now instead suppose M adds to A the constraint H: β≥0 (rather than β=0); then $R_M$ has the same dimension as $R_A$ and will contain interior points (i.e., points bounded away from $R_A - R_M$) within the ordinary induced topology of $R_A$. More generally, if the constraints in A beyond the distributional assumptions are defined using $k_A$ functionally independent equations, and H comprises $k_H$ more independent equations for a total of $k_M = k_H + k_A$ constraints in M, then $R_A$ and $R_M$ are $n-k_A$ and $n-k_M$ dimensional manifolds traced out in $R^n$ and $df_{M|A} = k_H$. But if instead $R_M$ adds only independent inequalities, $df_{M|A} = 0$ and $R_M$ will ordinarily have interior points relative to $R_A$.

**Divergence measures**

While there is no single formal definition of model regularity, under the usual definitions and models $R_A$ and $R_M$ will have the properties required for ease of analysis such as being closed and simply connected with smooth boundaries, as will be assumed here. Nonetheless, even with such regularity, divergence and NP-decision P-values can come into serious conflict when (as with interval H) some of the constraints in H are inequalities. The conflicts may be described as arising from differences in goals which lead to different treatments of $R_M$ when H contains inequalities or other problematic constraints.

NP theory aims to "test" M against A based solely on resampling criteria, making a decision about whether to reject M for poor fit relative to A (model checking) or equivalently whether to reject H given A (hypothesis testing). In contrast, divergence statistics aims to describe the relation of M to A in geometric terms by quantifying how much M diverges from A, according to some divergence measure between projections (images) of y on $R_M$ and $R_A$. These goals can be recast in terms of how much the information in y given M diverges from the information in y given A; or equivalently, how much information about F(y;μ) is lost when our estimate of it is constrained to the smaller distribution subspace ***F***$_M$ = {F(y;μ): μ ϵ R$_M$} instead of the larger subspace ***F***$_A$ = {F(y;μ): μ ϵ R$_A$}. As will be illustrated, technical details of information measurement, loss, and preservation via divergence measures can be represented via long-established concepts of loglikelihood information and its generalizations (Kullback, 1997; Amari, 2016).

Geometric and information concepts turn out to track each other closely in the setting considered here: Throughout most of practice if not theory, divergence is measured based on compatibility or information criteria that compare distributions in ***F***$_μ$ = {F(y;μ): μ ϵ R$_μ$} using transforms of summary statistics such as standardized distances in R$_μ$ or likelihood ratios in ***F***$_μ$. These lead to two major classes of divergence measures: Euclidean (squared-deviation) and information-theoretic (deviance) based. These classes approximate each other in common settings such as GLMs, and coincide in classical Gaussian linear models. Thus in typical large-sample settings they can be treated as different ways of viewing the same underlying discrepancies: The geometric measure expresses divergence in terms of differences in fitted data vectors in R$_μ$, while the information measure expresses divergence in terms of differences in information content (expected loglikelihoods) in ***F***$_μ$.

Let d(λ;θ) represent a divergence measure from F(y;θ) to F(y;λ) on ***F***$_μ$, with d(λ;θ) smooth and nonnegative over both λ and θ, and zero if and only if the compared distributions are equal almost everywhere; it may however be asymmetric in that d(θ;λ) is distinct from d(λ;θ). In the information-theory literature d(θ;λ) is more often written as D(λ‖θ). The present notation is adopted to parallel the logical ordering in F(y;μ) where μ is the index of the distribution being used to evaluate y, and also to avoid confusion with a random divergence which is denoted below using D. Information-divergence definitions may further specify that F(y;θ) is absolutely continuous with respect to F(y;λ); this restriction is however often left out of statistical discussions in favor of other devices such as setting 0·ln(0) to zero (e.g., see Bishop et al., 1975).

A geometric example is the F(y;θ)-standardized squared Euclidean distance SSD(λ;θ) = (λ−θ)′cov(Y;θ)$^{-1}$(λ−θ), or Fisher-information divergence, which is asymmetric if cov(Y;μ) varies over R$_μ$. An information-theoretic example is the deviance dev(θ;λ) = 2KLD(θ;λ) where KLD(θ;λ) is the Kullback-Leibler discrimination information, defined as the expectation over F(y;λ) of the log-likelihood ratio for F(y;λ) versus F(y;θ). KLD(θ;λ) is sometimes described as the expected information loss from using F(y;θ) as the distribution of Y when the true distribution is F(y;λ), or the entropy of F(y;θ) relative to F(y;λ).

Basic SSD and deviance measures will suffice to illustrate the present points, although there are many extensions and variations on them derived by marginalizing, conditioning or penalizing the distributions to improve some measure of statistical performance, ease computations, or introduce external (prior) information.

**Divergence statistics**

Suppose now we have selected a divergence measure d(y;μ) to efficiently capture information in the discrepancy of the observed data y from μ, scaled (standardized) by the corresponding distribution F(y;μ). The observed divergence d(y;R$_A$) of y from R$_A$ is then the infimum inf$_A$[d(y;a)] over R$_A$. If unique, the minimizer of d(y;m) over R$_A$, μ$_A$ = arginf$_A$[d(y;a)], is often taken as a point estimate of μ under the model A; μ$_A$ = y if y is an interior point of R$_A$ and will be a boundary point of R$_A$ otherwise. One may also use d(y;R$_A$) = d(y;μ$_A$) as a goodness of fit statistic for evaluating A against a mean-saturated alternative (for which y is the estimate of μ).

As an example, suppose d(y;μ) is the μ-standardized sum of squared deviations SSD(y;μ) which is the squared distance from μ to y under the metric (scaling) defined by cov(Y;μ); d(y;R$_A$) is then the minimum squared F(y;μ)-standardized distance to y over μ in R$_A$, and the closest point to y in R$_A$ is μ$_A$ = arginf$_A$[SSD(y;m)]. Furthermore, SSD(y;μ$_A$) is a score statistic for the fit of model A and reduces to Pearson's $\chi^2$ fit statistic. If cov(Y;μ) and hence the distance metric may vary with μ, iteration over μ in R$_A$ may be required to find d(y;R$_A$). The resulting SSD minimizer μ$_A$ is sometimes called an iteratively reweighted least-squares (IRLS) estimate of μ under A, and equals the maximum-likelihood estimate (MLE) when A is a family of ordinary GLMs; however, use of IRLS on a link-transformed deviation z−g(μ) instead of y−μ requires adjustment of z at each iteration (McCullagh & Nelder, 1989).

Further simplifications occur when using instead d(μ;y). For example, taking d(μ;y) = SSD(μ;y), the squared F(y;y)-standardized distance from y to μ, we get d(R$_A$;y) = inf$_A$[d(μ;y)] as the squared length of the orthogonal projection of y onto R$_A$ under the cov(Y;y) metric; μ$_A$ = arginf$_A$[SSD(μ;y)] is then a generalized least-squares (GLS) estimate of μ under A, leading to the Neyman $\chi^2$ statistic d(R$_A$;y) = SSD(μ$_A$;y). With standard fixed-dimensional GLMs for count Y, the Pearson and Neyman $\chi^2$ are first-order locally equivalent, since under A their covariance components converge to the same large-n limit; and for binomial-logistic and Poisson-loglinear models, they are score statistics using expected and observed information, respectively. Nonetheless, simulation studies and numerical considerations have suggested that the $\chi^2$ approximation to the fit statistic and the normal approximation to $\mu_A$ is better in small samples when using SSD(y;μ) and IRLS than when using SSD($\mu_A$;y) and GLS, thanks to the greater stability of the resulting cov(Y;$\mu_A$) compared to cov(Y;y) (Maldonado & Greenland, 1994). For example, in Poisson models cov(Y;μ) is undefined if any component of μ is zero, and typically $\mu_A$ will have zeros much less frequently than y. Taking instead d(μ;y) = dev(μ;y), d($\mu_A$;y) becomes the likelihood-ratio (LR) statistic for the fit of A, while $\mu_A$ = arginf$_A$[d(μ;y)] becomes the MLE of μ under A, which also appears to perform better than SSD($\mu_A$;y).

Consider now a submodel M of A. One measure of the divergence d($R_A$;$R_M$) of y from M given A takes the minimizer $\mu_A$ = arginf$_A$[d(y;a)] and then minimizes over $R_M$, so that
$$d(R_A;R_M) = \inf_M[d(\mu_A;m)] = d(\mu_A;\mu_M)$$
where $\mu_M$ = arginf$_M$[d($\mu_A$;m)] is the point estimate of μ under M given A; $\mu_M$ will be a boundary point of $R_M$ if $\mu_A$ is exterior to or on the boundary of $R_M$. We may call d($R_A$;$R_M$) a fit statistic for M given A, or a fit statistic for M against alternatives in A, or a test statistic for H = M−A given A. This case is also described by saying the data are filtered through the model A before being used to evaluate the submodel M.

Taking d($R_A$;$R_M$) = SSD($R_A$;$R_M$) we obtain a score statistic SSD($\mu_A$;$\mu_M$), where $\mu_A$ and $\mu_M$ are the IRLS estimates of μ, again equal to the MLEs of μ under A and M when both are ordinary GLMs. A reason for choosing F(d;$\mu_M$) for the standardization is that it makes $\mu_M$ the closest one can come to $\mu_A$ without leaving $R_M$ and thus violating H. Nonetheless, taking instead take d($R_M$;$R_A$) = dev($R_M$;$R_A$), $\mu_A$ = arginf$_A$[dev(a;y)] and $\mu_M$ = arginf$_M$[dev(m;$\mu_A$)] automatically become the MLEs of μ under A and under M, with dev($R_M$;$R_A$) = dev($\mu_M$;$\mu_A$) the deviance (likelihood-ratio) statistic for the fit of M given A.

The deviance and Neyman $\chi^2$ obey the additive (Pythagorean) relation d($\mu_M$;y) = d($\mu_A$;y) + d($\mu_M$;$\mu_A$), which reveals that M may appear to fit well or even perfectly given A and yet be a very poor fit to the actual data y, since we may have d($\mu_M$;$\mu_A$) ≈ 0 with d($\mu_M$;y) ≈ d($\mu_A$;y) arbitaritly large. The same caution applies to the Pearson $\chi^2$, although additivity does not hold exactly if cov(Y;μ) varies with μ (as with subset coherence, it may arise as a local asymptotic approximation which however breaks down when d($\mu_M$;y) is large). Nonetheless, numeric examples suggest that the usual $\chi^2$ approximations may produce less accurate P-values from dev($\mu_M$;$\mu_A$) and SSD($\mu_M$;$\mu_A$) than from SSD($\mu_A$;$\mu_M$), perhaps unsurprising given that those measures fail to use H when scaling the divergences, whereas SSD($\mu_A$;$\mu_M$) does use H.

**Parametric models**

Most often, A is a family of distributions indexed by a finite parameter vector β with parameter space ***B*** of dimension $k_A \ll n$ that remains fixed as n increases. The above concepts are then transformed into the much smaller, computationally more manageable space ***B*** in place of $R_\mu$. Elements of $R_A$ can then be written as μ(β), with a point estimate of β under A of $\beta_A$ = arginf$_B$\{d[y;μ(β)]\} and under M given A of $\beta_M$ = arginf$_B$\{d[μ($\beta_A$);μ(β)]\}. The resulting fit statistics d(y;$R_A$) = d[y;μ($\beta_A$)] and d($R_A$;$R_M$) = d[μ($\beta_A$);μ($\beta_M$)] measure the divergence of y from $R_A$ and of μ($\beta_A$) from $R_M$. Parallel notations follow when using instead d[μ(β);y]. The parameter-estimation choice then maps to the divergence choice, with IRLS corresponding to SSD[μ($\beta_A$);μ($\beta_M$)], GLS corresponding to SSD[μ($\beta_M$);μ($\beta_A$)], and ML corresponding to dev[μ($\beta_M$);μ($\beta_A$)].

Given the parametric structure of $R_A$, we can also define measures of divergence of observations from M given A in the parameter space ***B***. When evaluating a hypothesis H expressed as a linear constraint on β, most software uses a divergence in ***B*** in the form of the Wald statistic for H given A, defined as the $\beta_A$-standardized squared distance from $\beta_A$ to $\beta_M$,

$$\text{SSD}(\beta_M;\beta_A) = (\beta_A-\beta_M)'\text{cov}[B_A;\mu(\beta_A)]^{-1}(\beta_A-\beta_M)$$

where $B_A$ = arginf$_B$\{d[Y;μ(β)]\} is the random analog of $\beta_A$; using instead SSD($\beta_A$;$\beta_M$) corresponds to replacing the covariance matrix by cov[$B_A$;μ($\beta_M$)]. When β=μ and A is saturated, $\beta_A$ = $\mu_A$ = y and these statistics become the Neyman and Pearson $\chi^2$ statistics for the fit of M. More generally, under regular models both Wald statistics are first-order locally equivalent to the score SSD[μ($\beta_A$);μ($\beta_M$)] and deviance dev[μ($\beta_M$);μ($\beta_A$)]. Nonetheless, a long-standing literature on higher-order asymptotics (e.g., Efron & Hinkley, 1978) as well as simulation studies (e.g., Maldonado & Greenland, 1994) suggest that in finite samples the usual $\chi^2$ approximations for the Wald statistics are inferior to those for the other statistics.

**Divergence distributions**

Let $D_M$ be the random variable obtained by applying a divergence statistic d($R_A$;$R_M$) or d($R_M$;$R_A$) to repeated samples from F(y;μ). If $R_M$ contains more than one point, the distribution F(d;μ) of $D_M$ is unknown even if we are given that M is correct (μ ϵ $R_M$); hence it may be unclear what distribution for $D_M$ should be used to evaluate M. One common device is plug-in estimation, which takes μ = $\mu_M$ with adjustments for using the random model F(d;μ_M) in place of a fully prespecified model. An example is Fisher's reduction of the degrees of freedom for Pearson's $\chi^2$ statistic for M, SSD(Y;μ_M), from Pearson's original incorrect value to the residual $df_{M|A}$ = n−dim(R_M); here, A is saturated and hence dim(A) = n.

In other cases, deficiencies of the plug-in approach according to resampling criteria may lead to modifications of A or the model family *F*_A by finding suitable pivots or by marginalizing, conditioning, or penalizing the F(y;μ), thus altering the divergence comparisons in ways that may depend on the data, as well as requiring adjustments to degrees of freedom. These modifications can become difficult if H involves more than simple dimension reduction, as with inequalities or with penalty functions (even if the latter vary smoothly over the model space). Regardless of these difficulties, in the divergence approach the reference distribution serves only to provide a scale for evaluating the observed divergence d_M allowing for random error constrained as specified by the embedding model for d(R_M;R_A) or the target model for d(R_A;R_M).

**Compatibility P-values**

Divergence measures are difficult to interpret directly because, first, they correspond to squared distances rather than distances, and second, their size depends on many "nuisance" aspects of the problem including effective degrees of freedom. For example, one could replace the Pearson $\chi^2$ statistic SSD(y;μ_M) with its square root to make it a Euclidean distance in a space scaled by standard deviations, but its tail behavior and the identification of extreme values would still vary with specifics of M and A. Thus, to facilitate consistent interpretations, we transform each divergence into a P-value.

In the remainder of this review, I will use F(d_M;M) to denote a derived reference distribution for D_M. This will *not* be limited to a plug-in distribution, and may be a function of the data not only through μ_M but also through ancillary statistics or other data features; it may even have a Bayesian derivation, but is selected to meet frequency criteria (Bayarri & Berger, 1999, 2000, 2004). Let μ_t denote the unknown true value of μ. To avoid extensive technical complications, I will focus on large-sample behavior and assume F(d_M;M) converges to F(d_M;μ_t) as n increases when μ_t is in R_M, and does so at a rate sufficient for the proposed usage.

We can now define an observed approximate divergence P-value p_M for evaluating M as Pr(D_M≥d_M;M), the probability under F(d_M;M) that D_M would be at least as big as the observed value d_M of D_M; p_M =

1−$F(d_M;M)$ when $D_M$ is continuous at $d_M$. With H: $b_L \leq \beta \leq b_U$, $p_M$ reduces to the interval P-value described earlier. A key point is that $p_M$ is a descriptive statistic showing the ordinal location of the observed divergence $d_M$ in an explicit reference distribution whose choice may be dictated by information-summarization goals rather than decision goals. Confusion of these goals has fueled controversy over choice of sample space, as illustrated by the vast literature on whether we should condition on random margins of a 2x2 table (e.g., contrast Fisher, 1955 p. 70 with Shuster, 1992), with information summarization arguing for conditioning and decision criteria arguing against.

For descriptive purposes, $p_M$ can be treated as an index of compatibility of the data y with M given A. From a more precise geometric perspective, $p_M$ is a unit-scaled index of compatibility of the data image $\mu_A$ on $R_A$ with $R_M$, where compatibility is measured by a divergence scaled to the reference distribution $F(d_M;M)$, and information is measured either through inverse-covariance (information or precision) matrices or through log likelihoods. It should be noted that this perspective makes no reference to "inference", "significance", "confidence", or other subjective judgements about the contextual implications or importance of the observation. In particular, $p_M=1$ only means that $\mu_A$ is in $R_M$ and thus in this descriptive sense M is perfectly compatible with the data y given A, while ever smaller values of p correspond to ever larger divergence of $\mu_A$ from $R_M$ and thus ever more discrepancy between the data as fitted under the embedding model A and as fitted under the restricted model M.

Note that perfect compatibility with data given A does *not* mean that M is supported by the data, let alone correct or even plausible: Any other submodel M* of A with $\mu_A \in R_{M^*}$ will have $p_{M^*} = 1$ and thus will also be perfectly compatible with y given A by this measure, even if it otherwise conflicts with M. Suppose for example n=1, A says Y is normal($\mu$,1), M further says $\mu \leq 0$, M* further says $\mu \geq 0$, and y = $\mu_A$ = 0. Then both M and M* are perfectly compatible with y, in that $p_M = p_{M^*} = 1$; yet any prior for $\mu$ continuous at zero would assign zero probability to both models being true at once (M∩M*), leaving M and M* mutually exclusive almost everywhere.

Finally, like squared distance and the $\chi^2$ distribution from which $p_M$ is often derived, $p_M$ is "multi-tailed" or omnibus in that it does not distinguish directions of departure of y or $\mu_A$ from $R_M$.

**Properties of random divergence P-values**

To establish a reference distribution, we may ask what repeated-sampling properties can aid the interpretation of a single observation as an information summary rather than as a decision criterion. These properties include a type of conditional uniformity in the distribution of the random analog $P_M$ of $p_M$ when the true value $\mu_t$ is on the boundary of $R_M$, reflecting the fact that data generated from boundary points will more often appear ambivalent about whether M is correct than data generated from other points. This uniformity in divergence P-values is a consequence of limiting divergence behavior at boundary points, and leads to correspondence with NP P-values when M has no interior point; it is not however taken as a defining or even necessary feature of P-values.

In contrast, in NP theory uniformity is treated as a desirable or even defining property of P-values, as it ensures the test size (type-1 error rate over resampling) of the decision rule "reject if p≤α" does not exceed α, while maximizing power in typical settings. Conflict with divergence P-values thus arises when the distribution of $P_M$ is not approximately uniform under M. In particular, if $\mu_t \in R_M$ and $\Pr(\mu_A \in R_M; \mu_t)$ = $\Pr(D_M=0; \mu_t) > 0$, the distribution of $P_M$ will have a point mass at 1 and appear excessively conservative (dominating but not converging to uniformity). The mass can however be dealt with in descriptions by conditioning on whether $D_M>0$ or $P_M<1$.

To illustrate, consider a generalization of an interval hypothesis to the case in which M is the region between two parallel (nonintersecting) hyperplanes in $R_A$. If $\mu_t$ is exterior to $R_M$ (or equivalently, given $R_M$ is closed, if $\mu_t \in R_A-R_M$) then, as n increases, the probability of $\mu_A$ falling in $R_M$ approaches zero, $D_M$ will in probability increase without bound, and $P_M$ will converge to zero. On the other hand, if $\mu_t$ is interior to $R_M$ (i.e., bounded away from $R_A-R_M$) the probability of $\mu_A$ falling in $R_M$ will approach 1 and so $P_M$ will converge to 1. These properties seem natural and desirable: Given a linear boundary, the standardized or information distance between $\mu_A$ and $R_M$ should increase with n if $\mu_t$ is outside of $R_M$, and should go to zero with increasing n if $\mu_t$ is inside $R_M$.

Suppose however $\mu_t$ is on a linear boundary segment of $R_M$ and bounded away from other segments, as in the parallel case. Then Y will typically have limiting nonzero probabilities of falling outside and inside of $R_M$, and no amount of data can lead to Y consistently falling in or out of $R_M$. When n is large enough so that the divergence statistic concentrates away from any other segment, the two probabilities will approach 50% as n further increases, so that the masses $\Pr(D_M=0; \mu_t)$ and $\Pr(P_M=1; \mu_t)$ will converge to 0.50. Thus, again assuming our reference distribution $F(d_M;M)$ converges to $F(d_M;\mu_t)$, the distribution of $P_M$ will develop a spike of height 0.50 at $p_M=1$ as n increases, while the remaining 50% of its mass will approach uniformity.

Outside of such examples, a subtle point for proper interpretation of $p_M$ is that if $R_M$ is convex and $\mu_t$ is on the boundary of $R_M$ but that boundary is not linear within most of the mass of $F(y;\mu_t)$, $Pr(D_M=0;\mu_t)$ can be much less than 0.50; this happens in the interval case when $\beta_t = b_L$ and there is high probability of $\beta_M > b_U$ (making y exterior to $R_M$ on the side opposite $\mu_t$). Note also that if $\dim(R_M)<\dim(R_A)$, in regular models $Pr(D_M=0;\mu_t)$ will go to zero as n increases and hence when M holds the data will almost always display some positive divergence from $R_M$. Indeed, for $\dim(R_A)-\dim(R_M) \geq 3$ the mass of a $\chi^2$ divergence statistic will be concentrated almost entirely away from 0 even if $\mu_t \in R_M$. This fact was the basis for Fisher's (1936, p. 130) renown conclusion that Mendel's genetic data was biased, where the presumed violation of M was underdispersion relative to the assumed data distribution (suggested by the suspiciously close fit to expectations) rather than departure from expectations.

**Interpreting P-values via S-values (surprisals)**

Unconditionally, smaller $p_M$ correspond to greater incompatibility of y with M given A, according to the reference distribution for $D_M$. If $p_M = 1$ we know $\mu_A$ is in $R_M$, which is to say M is perfectly compatible with y given A; otherwise, if $p_M < 1$, we know $\mu_A$ is outside of $R_M$, $D_M > 0$, and $\mu_M$ is a boundary point, with smaller p corresponding to greater standardized distance between $\mu_A$ and $\mu_M$.

While $p_M$ at first appears convenient as a 0-to-1 index of compatibility, it is poorly scaled in that, with typical distributions, small $p_M$ differences near 1 represent trivial divergence differences whereas small $p_M$ differences near 0 represent dramatic divergence differences. This observation leads to rescaling $p_M$ with a negative log transform. As an example, the binary S-value or surprisal at seeing $\mu_A$ given M, $s_M = \log_2(1/p_M) = -\log_2(p_M)$, can be used as a measure of the information against M supplied by the observed divergence. Specifically, if whenever M holds, $P_M$ stochastically dominates a unit-uniform variate, $s_M$ is a lower bound on the Shannon information in $p_M$ against M given A, expressed in log base 2 units (bits), or equivalently a lower bound on the number of bits of statistical information against M supplied by the event $D_M=d_M$, which is within 1 bit of the information when $Pr(D_M=0;\mu_M) \leq 0.50$, with equality when P is unit uniform. Rafi & Greenland (2000) illustrate how $s_M$ varies with likelihood ratios and deviance statistics.

We can also compare the binary S-value to the information in a coin-tossing experiment about the hypothesis H: "the tossing is not biased toward heads": With $s_M$ rounded to the nearest integer $s_N$, the information against M supplied by the observed divergence $d_M$ is (within a half bit) the same as the information against H supplied by seeing all heads in $s_N$ tosses of a coin (Greenland 2019a; Rafi & Greenland 2020; Cole et al. 2021; Amrhein & Greenland, 2022; Greenland et al. 2022). Note again however that for more complex settings, *absence* of information *against* H need not correspond to *presence* of supporting information *for* H if the apparent absence could be due to violations of auxiliary assumptions in A = M−H (Greenland & Rafi, 2021; Bickel 2022).

**Compatibility regions**

Suppose now that within A we have a family of models $M_\beta$ indexed by a parameter β (which may be a vector) in a space ***B***, each defining a subset $R_{M\beta}$ of $R_\mu$ with its own divergence $d_{M\beta}$ and an accompanying reference distribution $F(d;\mu_{M\beta})$ and P-value $p_{M\beta}$. Generalizing the notion of compatibility intervals (Amrhein et al., 2019; Rafi & Greenland, 2020; Cole et al., 2021; Amrhein & Greenland, 2022; Greenland et al. 2022, 2023), we may then define a π-compatibility set for β as the set of all β for which $p_{M\beta} > \pi$. Usually this set will be a connected region in ***B***, and so can be called an π-compatibility region for β, or π-compatibility interval if β is a scalar. The use of π rather than α is to avoid confusion with the usual 1−α coverage or "confidence" region, the latter being equivalent to the region in ***B*** in which an α-level NP test fails to reject $M_\beta$.

When A is correct, the rule "reject $M_\beta$ if $p_{M\beta} \leq \alpha$" provides a valid α-level test for all β, and thus the corresponding α-compatibility region will be a 1−α coverage region as well. As with divergence P-values, however, compatibility regions are not optimized according to NP criteria (e.g., for uniformly most accurate symmetric coverage), and so more generally the two types of regions may fail to coincide, with compatibility regions being conservative in the NP sense of overcoverage. Paralleling arguments for divergence P-values, in such cases one may see that a compatibility interval conveys more information about what was actually observed than do realized NP decision ("confidence") intervals. The failings of observed NP decision intervals are unsurprising given that they are outputs of algorithms optimized only for resampling criteria, without regard to individual sample anomalies such as empty intervals (which have led some authors to reject frequentist methods entirely).

**Multiple comparisons**

The divergence view offers a relatively simple approach to multiple comparisons when the latter can be expressed in terms of a multidimensional constraint in a model. To illustrate, suppose the individual observations are row vectors $y_i = (y_{i1},…,y_{iJ})$ and the embedding model A states the corresponding $Y_i$ are independent identical J-variate normal row vectors with unknown mean vector μ and known covariance matrix Σ. Then the classic Scheffe'-type simultaneous (multiple-comparison) P-value for H: μ=0 (or more precisely, for M = H ∪ A given A) reduces to the percentile location (tail area) of a J df $χ^2$ (score) statistic comparing the observed-mean vector $\bar{y}$ to the zero vector; more precisely, this statistic is the squared Σ-standardized distance in the J-space $R_A$ from the origin **0** to $\bar{y}$. Here, subspace $R_M$ contains only the origin.

The simultaneous P-value should be contrasted to the J separate P-values $p_1,…, p_J$ for the single-component hypotheses $H_j$: $μ_j = 0$. Each $p_j$ is a percentile location of a 1 df $χ^2$ statistic comparing one component mean $\bar{y}_j$ to 0; that statistic is the squared standardized distance to $\bar{y}_j$ from the J−1 dimensional subspace $R_{Mj}$ defined by $H_j$. Viewed in this geometric fashion, we may see that, *whether or not one is interested in multiple comparisons and error control over multiple hypotheses*, the simultaneous P-value and the single-component P-values are describing relations among very different data summaries and subspaces. Only the research question can determine which (if any) of these or other P-values should be presented – a point that has been made repeatedly in purely logical terms (e.g., Greenland & Hofman, 2019; Greenland, 2021b; Rubin, 2021). If mapped carefully into the research context, geometric descriptions clarify the difference among the targets of multiple and single comparisons.

**Relations to α-level decision P-values**

Under the regular GLM set-up assumed here, suppose H is a pair of linear inequalities that make the boundaries of $R_M$ parallel hyperplanes (as when H is an interval). $Pr(P_M≤α;μ_t)$ then depends only on precision in the direction orthogonal to the two boundary hyperplanes. In large samples with $μ_t$ on one of these boundary segments, $Pr(D_M>0;μ_M)$ will approach 0.50; thus, with $P_M$ uniform given $D_M>0$, $Pr(P_M≤α;μ_t)$ will approach α/2. This suggests a bound on the inefficiency or information loss and consequent test conservatism from using $p_M$ to form a decision rule "Reject H if $p_M≤α$".

Suppose we have a decision-optimized P-value $P_T$ that is uniform under M. Upon observing $P_T = p_T$ we may take $s_T = −\log_2(p_T)$ as an upper bound on the bits of information against H given A, and compare this to the lower bound provided by $s_M = -\log_2(p_M)$. When $\mu_t$ is interior to $R_M$, both $P_T$ and $P_M$ approach 1 as n grows and so the gap between these information bounds approaches zero. When $\mu_t$ is exterior to $R_M$, the ratio $P_M/P_T$ approaches 2 as n grows, corresponding to a one-bit difference in Shannon information against M given A.

**Summary and Conclusion**

When M has no interior point relative to the embedding model A, as when H = M−A corresponds to dimension reduction with $R_A$ and $R_M$ proper vector subspaces of $R_\mu$, resampling-decision (NP) criteria and information-summarization (divergence) criteria may lead to an identical observed P-value p (modulo small-sample differences, as in unconditional vs. conditional tests in 2x2 tables). This numeric identity arises in most applications and may well explain why few statistics texts distinguish the two types of P-values. Nonetheless, they will differ in important examples of interval H; in those cases the question of which (if either) is more relevant will arise. A simple answer is that if the analyst is only trying to summarize model fits to the data in hand and wishes to focus on coherent measures of fit, the divergence P-value is the more relevant: It is a quantile transform of a divergence measure, and as such automatically meets the coherence criterion that the P-value for a nested model cannot exceed the P-values for its embedding model when, for example, it uses the the deviance as the measure.

As seen in simple examples where $R_M$ has interior points, a divergence P-value will sacrifice repeated-sampling optimality to maintain single-sample coherency. But those pure repeated-sampling criteria have no bearing on its interpretation as a geometric information summary: $p_M=1$ tells us the image $\mu_A$ of y in the embedding model A is also in M, while a very small $p_M$ indicates $\mu_A$ is stochastically far from M. In this manner, it measures how compatible y is with M under an embedding model, where $p_M=1$ means zero divergence and hence complete compatibility, with more divergence and hence lower compatibility indicated by smaller $p_M$. This compatibility interpretation is logically weaker than support, and is consistent with the conclusion of Schervish (1996) and others that P-values are poor measures of support, regardless of the definition of "P-value".